%% file: main.tex
\documentclass[useAMS,usenatbib,a4paper]{mnras}
 \usepackage{epsfig}
 \usepackage{relsize}
 \usepackage{josty}
 \usepackage{amsmath}
 \usepackage{animate}
 \usepackage{graphicx}

\def\sigtwo{\Sigma_{\rm *,2 kpc}}
\def\sign{\Sigma_{\rm *,N kpc}}
\def\bhmdot{\dot{M}_{\rm BH}}

\title[Structural Evolution at Low-$z$ in IllustrisTNG] 
{The Structural Evolution of Isolated Galaxies at Low Redshift in {the} IllustrisTNG {Simulation}}
\author[D. Walters et al.] 
    {\parbox{\textwidth}{Dan Walters,$^{1}$\thanks{danwalters@uvic.ca}
    Joanna Woo,$^{2,1}$
    Sara L. Ellison,$^{1}$
    and Maan H. Hani$^{3,1}$\textsuperscript{\thanks{Herschel Fellow}}} 
\vspace{0.4cm}\\
\parbox{\textwidth}{ 
$^{1}$Department of Physics \& Astronomy, University of Victoria, PO Box 1700 STN CSC, Victoria BC V8W 2Y2, Canada\\
$^{2}$Department of Physics, Simon Fraser University, 8888 University Drive, Burnaby BC V5A 1S6, Canada\\
$^{3}$Department of Physics \& Astronomy, McMaster University, 1280 Main Street West, Hamilton ON L8S 4M1, Canada
}}

\pagerange{\pageref{firstpage}--\pageref{lastpage}} \pubyear{2019}

\hypersetup{draft}
 
\begin{document}

\include{illustris_paper}

\include{appendix}

\end{document}

%% file: illustris_paper.tex
\label{firstpage}
\maketitle
 
\begin{abstract}
We study the structural evolution of isolated star-forming galaxies in the Illustris TNG100-1 hydrodynamical simulation, with a focus on investigating the growth of the {central} core density {within 2 kpc} ({$\sigtwo$}) in relation to total stellar mass ($\Ms$) at $z<0.5$.  
First, we show that several observational trends in the $\sigtwo$-$\Ms$ plane are {{qualitatively}} reproduced in IllustrisTNG, including the distributions of AGN, star forming galaxies, quiescent galaxies, and radial profiles of stellar age, sSFR, and metallicity. We find that galaxies with dense cores evolve parallel to the $\sigtwo$-$\Ms$ relation, {while} galaxies with diffuse cores evolve along shallower trajectories. {We investigate} {possible drivers} of rapid growth in $\sigtwo$ compared to $\Ms$. {Both the current sSFR gradient and the BH accretion rate are indicators of past core growth, but are not predictors of future core growth.}
Major mergers {(although} rare in our sample{; $\sim$10\%)} cause steeper core growth, {{except for high mass ($\Ms \gtsima 10^{10} \Msun$) mergers, which are mostly dry.}} 
Disc instabilities, as measured by the fraction of mass with Toomre $Q < 2$, are not predictive of rapid core growth.  Instead, rapid core growth results in more stable discs.  
{The cumulative black hole feedback history} sets the maximum {rate of} core growth, preventing rapid growth in high-mass galaxies ($\gtsima 10^{9.5} \Msun$). For massive galaxies the total specific angular momentum of accreting gas is the most important predictor of future core growth. Our results suggest that the angular momentum of accreting gas {controls} the slope, width and zero-point evolution of the $\sigtwo$-$\Ms$ relation.
\end{abstract}


\section{Introduction}
\label{introduction}

{The structural evolution of galaxies is an integral part of {their} overall evolution.  The sizes, morphologies and {central stellar mass} densities of galaxies correlate with their {total stellar} masses, star formation activity and environments \citep{Strateva2001,Kauffmann2004,Bell2008,Wuyts2011,Mendel2013,Schawinski2014,Bluck2014}. The average galaxy size at fixed stellar mass {grows} with {decreasing redshift} \citep{VanDerWel2014,Barro2017} in part due to the growth of individual galaxies \citep{Hopkins2009,Naab2009,Damjanov2011,vanDokkum2014}.  Furthermore, low star formation activity (quiescence) {correlates} more strongly with various measures of size/morphology than with any other galaxy property (the ``morphology-quiescence'' relation - \citealp{Woo2019}; see also \citealp{Bell2008,Bluck2014,Woo2015,Teimoorinia2016}).}

The basic physical framework for {galaxy} growth {is} mostly understood: {galaxies are expected to build up their stellar mass from the inside-out as a consequence of the top-hat spherical collapse model where expanding shells of dark matter turn-around and collapse forming dark matter halos.} {The innermost shells collapse first, as a result, the} gas with the lowest angular momentum {settles into} the galaxy earliest {which gives rise to negative} age gradients (the central regions are the oldest; \citealp{Fall1980,Kepner1999,vandenBosch1998,vandenBosch2002}).  This inside-out growth of galaxies is confirmed in hydrodynamical simulations (e.g., \citealp{Roskar2008,Tissera2016,Tacchella2016a}), and inferred in observations \citep{Nelson2012,Perez2013,Tacchella2015a}.  {The} inside-out {growth} mode can be thought of as the ``default'' mode for galaxy growth.  

Deviations from the inside-out mode include dramatic events such as high-$z$ ``compaction'' that drives gas (and sometimes stars) from the outside-in to the central regions of galaxies.  Such events produce a central starburst that significantly increases the central stellar density (or perhaps equivalently, grows the bulge) rapidly \citep{Zolotov2015}.  Dissipative inflows of gas can be triggered by galaxy mergers \citep{Barnes1991,Mihos1996,Hopkins2006} or gravitational disc instabilities \citep{Friedli1995,Immeli2004,Bournaud2011}.  These inflows build the central density causing steeper density profiles. The steeper density profiles result in smaller effective radii (hence ``compaction'').  Although events such as compaction appear to be deviations from the default inside-out growing mode, there are indications that compaction is an important event in the life of most galaxies, occuring primarily around $z\sim 2$ (Dekel et al., in preparation).  

{
In addition to increasing the central density of a galaxy, compaction events also potentially feed active galactic nuclei (AGN), which are thought to be necessary for quenching galaxies (eg. \citealp{Cattaneo2007,Woo2015,Nelson2018}).  The association of AGNs with the growth of the central density is one proposed explanation for the morphology-quiescence relation \citep{Zolotov2015,Woo2019}.  
}

{
Although compaction was originally envisioned as a high-$z$ phenomenon, \cite{Woo2019} (hereafter WE19) found evidence at $z < 0.07$ for both the default inside-out growth mode as well as a ``compaction-like'' mode.  They studied gradients of stellar age and star formation rate (SFR) in the Mapping Nearby Galaxies at the Apache Point Observatory survey (MaNGA - \citealp{Bundy2015}) and found that while galaxies with low central densities have older centres and lower central specific SFRs (sSFR) than their outskirts, galaxies with higher central densities tend to have relatively younger centres and more centrally-concentrated star-formation.  WE19 interpreted the latter observation as a low-$z$ version of ``compaction'' (listing various triggers), but acknowledged other possible interpretations. 
}

{
\cite{Tacchella2019} studied the morphological evolution of galaxies during quenching in the IllustrisTNG cosmological simulation \citep{Nelson2018,Pillepich2018,Springel2018,Naiman2018,Marinacci2018,Nelson2019}.  
Although they point out that the growth of the SMBH is related to the growth of the bulge, they {argue} that the dominant reason for the morphology-quiescence relation is the early quenching of most of today's quiescent galaxies, when concentrations were high and spheroids form most efficiently.  Yet the analysis of \cite{Tacchella2019} {also} shows that galaxy concentrations do increase during the transition to quiescence (see their Fig. 9), and therefore morphological transformation during quenching cannot be ruled out.  
It remains unclear whether the increase in central density (or concentration or bulge mass) during quenching is a significant transformation compared to the growth of the stellar mass.
}

{The question of morphological/structural transformation during quenching is not a trivial matter, and requires that we first understand the structural growth of {\it star-forming} galaxies.  Therefore, the aim of the current study is to investigate the growth of the central density in isolated star-forming galaxies.} {We investigate the growth of galaxy central density} in the IllustrisTNG simulation.  We first check that the age and sSFR gradients vary with central density and stellar mass at $z\sim 0$ in the same qualitative way as observed in the real universe (they do).  Then, we investigate the physical mechanisms that give rise to these gradients.  
In particular, why do galaxies with high central densities have centrally-concentrated sSFRs?  Are compaction-like processes responsible for their high central densities, or were their high central densities established long ago?  WE19 argue that compaction-like events will result in steeper evolutionary paths in the $\sigone$-$\Ms$ diagram compared to the inside-out mode of galaxy growth (see also \citealp{Barro2017,Zolotov2015}).  Does the steepness of these paths correlate with certain galaxy properties that lead to the build-up of the central density?  Among these properties, we investigate the role of disc instabilities, the occurance of mergers, black hole feedback, and the angular momentum of the accreting gas as possible reasons for the build-up of dense cores.

{{This paper is structured as follows. First, in Section \ref{methods} we discuss our methodology. Next, in  Section \ref{results} we present our results, including a comparison of TNG with observations, an examination of the evolutionary pathways of TNG subhalos, and an investigation of possible drivers of core growth. In Section \ref{discussion}, we discuss the implications of our findings. Lastly, we summarize our conclusions in Section \ref{summary}.}}

\section{Methods}
\label{methods}

\subsection{Simulation}
\label{simulation}

Illustris TNG (hereafter, TNG) is a set of large-box, cosmological gravo-magnetohydrodynamical simulations evolved from $z$ = 127 to the present \citep{Nelson2018,Pillepich2018,Springel2018,Naiman2018,Marinacci2018,Nelson2019}. TNG assumes a $\Lambda$CDM cosmology with parameters consistent with the Planck 2015 results \citep{Ade2016}. The physical model employed in TNG includes a variety of subgrid physics, including: stochastic star formation, supernova feedback, stellar evolution and winds, and supermassive black hole seeding, growth, and feedback (including both low- and high- accretion rate modes). The TNG model is calibrated to reproduce certain galaxy statistical properties, including: the size-mass relation, the cosmic star formation rate (SFR) {density}, and the black-hole (BH)-stellar mass relation \citep{Pillepich2018}. Halos and subhaloes (galaxies) are identified using {friends-of-friends} (FOF) and SUBFIND algorithms respectively \citep{Springel2001}. Note that we use the terms subhalo and galaxy interchangeably. 

AGNs in TNG are modeled to inject energy into the surrounding gas in either a low- or high-accretion rate mode \citep{Weinberger2017}. The low accretion rate mode injects kinetic energy, while the high accretion rate mode injects thermal energy. The mode each BH is in depends primarily on the accretion rate and the mass of the BH, where the scaling relation favors the thermal mode for low-mass ($\ltsima 10^{8.2} \Msun$) black holes, and the kinetic mode for high-mass black holes \citep{Terrazas2020}. The kinetic mode couples much more efficiently with the surrounding gas, and lengthens the cooling time of the circumgalactic medium \citep{Zinger2020}. In either case, the energy released is directly proportional to the accretion rate.

In this paper we use the TNG100-1 simulation which is the highest resolution volume that is currently publicly available. TNG100-1 simulates a cosmological cube with a comoving side length of 100 cMpc. In TNG100-1 there are 22554 subhaloes resolved with $\Ms > 10^9 \Msun$ at z = 0. In order to check the effects of spatial resolution, we also analyse the lower resolution version TNG100-2. This is discussed in {Section} \ref{caveats}.

\subsection{Subhalo Selection}
\label{selection}

We limit our analysis to galaxies whose z = 0 descendant is an isolated central galaxy with $\Ms > 10^9$ $\Msun$. We define a central subhalo as isolated if the second most-massive subhalo in the FOF group has a stellar mass $<$ 20\% of the central's stellar mass, and $<10^9$ $\Msun$. Lastly, we remove any subhaloes flagged by the TNG collaboration as non-physical.

We further divide our sample into star forming (SF), quiescent (Q), and a third category we dub ``intermediate'' galaxies. {{In the simulation, subhaloes with SFR below the numerical resolution limit are assigned SFR $=0$ (see \cite{Donnari2019} for a discussion).}}  Quiescent subhaloes have {unresolved SFRs (i.e. SFR $=0$)}, while SF subhaloes have sSFR $>$ $10^{-11} $yr$^{-1}$ which corresponds to the commonly used observational limit of SF vs Q galaxies. Observationally, {the non-zero sSFR values reported for quiescent galaxies are actually upper limits.}  We define subhaloes with 0 $<$ sSFR $\leq$ $10^{-11} $yr$^{-1}$ as ``intermediate'' galaxies. 

Our goal is to investigate ``low-redshift'' evolution in order compare our results with \cite{Woo2019}, therefore we investigate TNG galaxies at z $<$ 0.5 ($\sim$5 Gyr in lookback time) which gives us sufficient history to determine galaxy evolutionary trends. The snapshot time resolution of our sample is $\sim$113 to 236 Myr. 

\begin{figure}
\includegraphics[width=0.9\linewidth]{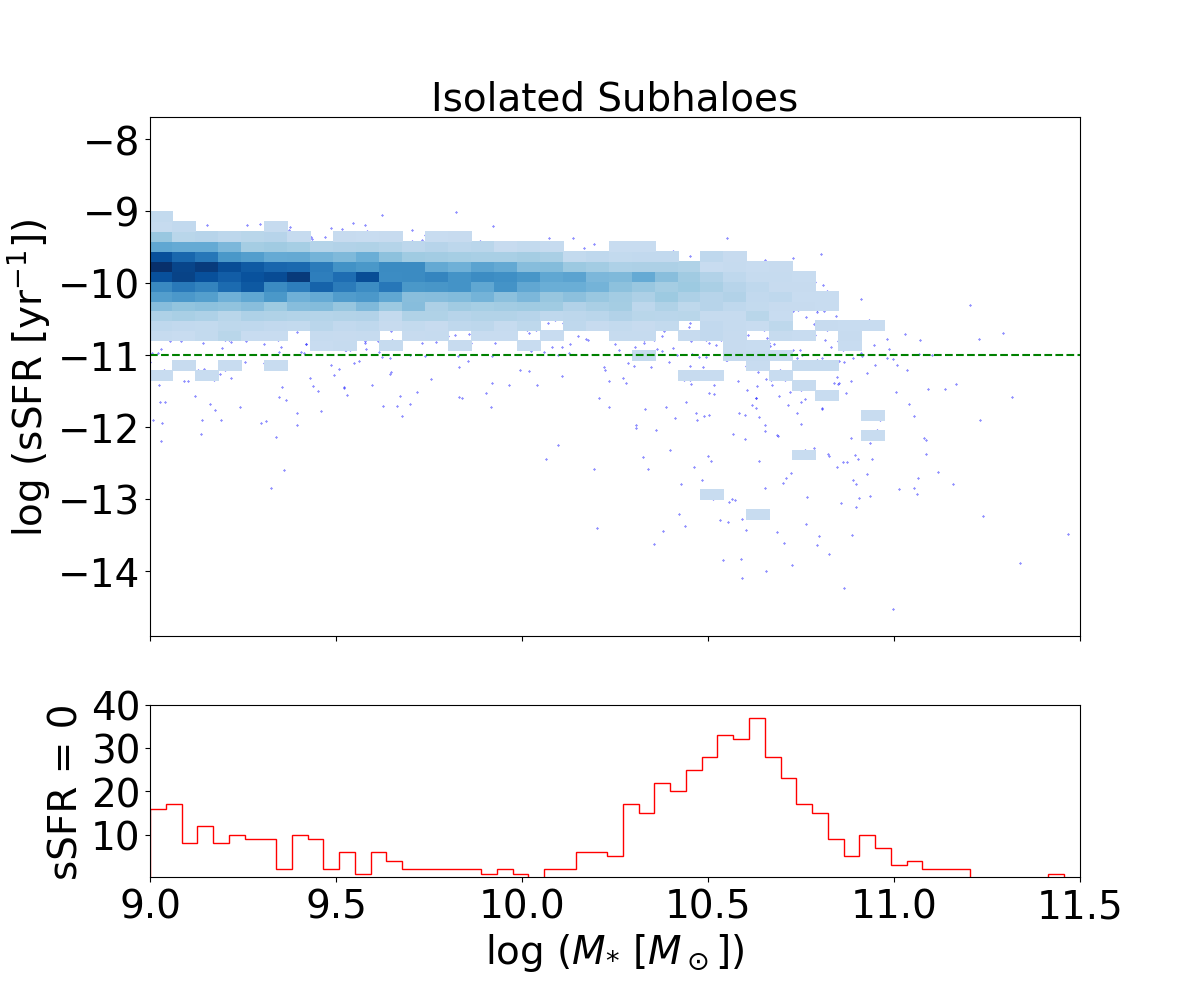}
\caption{\small Top panel: The sSFR-$\Ms$ relation in TNG at $z=0$. The horizontal dashed line shows the limiting sSFR required for subhaloes to be defined as star-forming. Below this dashed line, subhaloes are classified as 'intermediate' if the 0 $<$ sSFR $\leq$ $10^{-11} yr^{-1}$ and quenched if sSFR = 0. Lower panel: Distribution stellar mass in the subhaloes with unresolved SFR (i.e. quenched).  } 
\label{trans}
\end{figure}

\subsection{The Central {Stellar Mass Surface} Density {$\sigtwo$}}
\label{sigcalc}

{Our goal is to investigate the coevolution of central stellar density and stellar mass. The central density is commonly measured as the stellar mass surface density within a fixed radius, e.g., 1 kpc ($\sigone$) or 2 kpc ($\sigtwo$), or generally $\sign$, to represent the ``compactness'' of a galaxy's core  \citep{Cheung2012,Fang2013,Woo2015,Mosleh2017,Tacchella2017,Woo2017,Woo2019}. {Simulations show that }fixed-radius surface densities remain remarkably constant when star-formation ceases in the core, whereas densities within a scale radius (eg. the half-mass radius) experience large fluctuations due also to changes in the galaxy's outskirts \citep{Zolotov2015,Wellons2016,Barro2017}. In the absence of new star-formation, $\sign$ decreases slightly due to stellar mass loss as the population ages, but overall, the primary changes to $\sign$ are driven by dissipative gas processes such as compaction, making $\sign$ a good tracer of such processes.
} 
In any case, for undisturbed star-forming galaxies, $\sign$ tightly correlates with the effective radius \citep{Luo2020,Chen2020}. 

In this work, we use 2 kpc to quantify the central density ($\sigtwo$), so as to sit comfortably above the resolution of TNG (gravitational softening length 0.74 kpc). {The surface density within a fixed radius is subject to projection effects, largely ignored in observational data.  To investigate these effects, we measure {$\sigtwo$} for TNG galaxies using three methods. }
First, we project all subhalo star particles onto a random plane, then sum the mass of particles within the desired radius of the subhalo centre. Second, we project the star particles onto a plane perpendicular to the subhalo stellar component angular momentum vector before summing, giving a ``face-on'' measure of {$\sigtwo$}.  Third, we sum the mass of star particles within the 3-D desired radius from the subhalo centre (i.e. without projecting). In all cases, we divide by the surface area, $\pi r^2$. We note that in the case of the 3-D value, the calculated $\sigtwo$ is not a true surface density. However, it is useful for investigating whether any {``observed'' compaction-like behaviour} is biased by projection effects rather than a true core density increase. 
{We find that projection effects tend to artificially increase {$\sigtwo$} by $\sim$0.2 dex for spheroids and $\sim$0.3 dex for disc galaxies compared to the true 3-D mass within 1 kpc divided by $\pi r^2$.  Since these effects are independent of mass, the overall effect is a vertical shift in the {$\sigtwo$-$\Ms$} relation that is slightly different between quiescent and star-forming galaxies.}  
When examining subhalo histories we use the face-on projected value to remove any possible effects of projection should the subhalo's angular momentum direction change over time. When comparing to observations, we use the random projection to give as close a comparison as possible. 

{Although $\sigone$ pushes the resolution of TNG100-1, we repeated our analysis using $\sigone$ and find essentially the same trends as with $\sigtwo$.}

\subsection{Calculation of Radial Gradients}
\label{radgrad}

{
In order to verify that TNG reproduces observed gradients as a function of {$\sigtwo$} and $\Ms$ (WE19), 
}
we calculate stellar age, sSFR, and gas {phase} metallicity gradients for all TNG subhaloes. We compute ``two-point'' radial gradients using quantities measured within the stellar half mass radius ($R_e$) and $2R_e$. While this is not strictly analogous to the method used by WE19, our method provides a qualitative comparison with WE19. As we shall see in \secref{results}, the two-point radial gradient in stellar age and sSFR computed for the TNG galaxy sample are in qualitative agreement with the WE19 observational results. 

In the case of stellar age, we calculate the mass-weighted average age of all star particles within the area of interest as follows: 

\begin{equation}
  \nabla_r \log {\rm age} = \frac{\log \frac{\sum\limits_{\rm out} (m_*)(\rm age_*)}{\sum\limits_{\rm out}m_* } - \log\frac{\sum\limits_{\rm in} (m_*)(\rm age_*)}{\sum\limits_{\rm in}m_* }}{\log(1.5R_e) - \log(0.5R_e) }
\end{equation}

For sSFR and gas phase metallicity, we calculate gradients as:

\begin{equation}
 \nabla_r {\rm sSFR} = \frac{\log\frac{\sum\limits_{\rm out} {\rm SFR}}{\sum\limits_{\rm out} m_*} - \log\frac{\sum\limits_{\rm in} {\rm SFR}}{\sum\limits_{\rm in} m_*}}{\log(1.5R_e) - \log(0.5R_e) }
\end{equation}
\begin{equation}
 \nabla_r \log {\rm O/H} = \frac{\log\frac{\sum\limits_{\rm out} m_{\rm gas,O}}{\sum\limits_{\rm out} m_{\rm gas,H}} -\log\frac{\sum\limits_{\rm in} m_{\rm gas,O}}{\sum\limits_{\rm in} m_{\rm gas,H}}}{\log(1.5R_e) - \log(0.5R_e) }
\end{equation}

The subscript ``in'' refers to particles within $r<R_e$ while ``out'' refers to particles within $R_e<r<2R_e$.

\subsection{Tracking Galaxy History}
\label{trackhist}

We track subhaloes' histories using the SubLink merger trees \citep{Rodriguez-Gomez2015}. For each isolated subhalo selected at $z = 0$, we track the main progenitor branch (MPB) back to $z = 0.5$ ($\sim$ 5.2 Gyr). We choose $z = 0.5$ as it represents the relatively recent universe, while providing sufficient evolutionary histories for the galaxies within our sample.

{We tag each subhalo's history any time a} progenitor on {its} MPB undergoes a major merger. 
{Following \cite{Rodriguez-Gomez2015}, we define a major merger as having a mass ratio of at least 1:10 between the subhalo's most massive and next most massive progenitors. }
To avoid histories that undergo subhalo ``switching'', we follow the method used by \cite{Genel2017}. Specifically, we exclude from our analysis subhaloes whose MPB experiences a relative mass drop of $>$ 0.5 over one snapshot.

\subsection{Calculation of Toomre Parameter}
\label{toomre}

As part of our analysis, we calculate the Toomre stability parameter (Q) of the gas and stellar disc components of the TNG subhaloes. We roughly follow the method used by \cite{Inoue2015}. The Toomre parameter $Q$ measures the local stability in response to an axisymmetric density perturbation for a thin disc \citep{Toomre1964}, and it is given by:

\begin{equation}
Q = \frac{\sigma\kappa}{AG\Sigma}
\end{equation}
where $\sigma$ is the velocity dispersion, $\kappa$ is the epicyclic frequency (a measure of the potential), A is a constant ($\pi$ for the gas component or 3.36 for the stellar component), G is the gravitational constant, and $\Sigma$ is the surface density. All values are local values, as is $Q$. Areas with $Q < Q_{crit}$ indicate local instability. $Q_{crit}$ is classically 1, although discs may be unstable up to $Q \sim 2$ \citep{Elmegreen2011}. Initially, we compute $Q$ separately for the stellar and gaseous disc components, then combine them to create a single $Q$ map for each subhalo. 

We smooth all physical values by treating the simulation particles as Gaussian functions with fixed smoothing length. The results presented here use a smoothing length of 0.74 kpc, equivalent to the gravitational softening length for collisionless particles in TNG100-1, and therefore the maximum resolution we can achieve for resolving instabilities.  To combine values into our $Q$ maps, we apply a mesh to each subhalo with a fixed number of cells within $2R_e$ (i.e. over a box $4R_e$ x $4R_e$). Physical values are calculated at the centre of each mesh point. The results here use a 70 x 70 mesh. Prior to computing Toomre maps at all snapshots using our fiducial smoothing length and mesh size, we tried varying both for a single snapshot. {We find no qualitative difference in single snapshot distribution if we vary the smoothing length as low as 0.37 kpc nor if we vary the mesh to as fine as 140 x 140 within the same box}.

{Our first step in creating maps of Q for each galaxy is to determine the density of the stellar and gas disc components.} To separate the stellar disc and bulge components we follow \cite{Inoue2015}. We calculate the angular momentum component of each star particle parallel to the angular momentum of the subhalo ($J_z$) and compare it to the angular momentum of a circular orbit with the same energy ($J_c$). Star particles with $J_z/J_c < 0.7$ are tagged as bulge stars, and are excluded from our analysis (except for their contribution to the potential and therefore to $\kappa$). For gas particles, following \cite{Inoue2015}, we include all gas particles within 3 kpc of the disc plane.

Next, we calculate the radial velocity dispersion of the gas and stellar components. For the gas component, we first calculate the local speed of sound ($c_s$) and the radial velocity dispersion ($\sigma_{t}$), then calculate the combined $\sigma^{2}_{gas} = c^{2}_{s} + \sigma^{2}_{t}$. As noted by \cite{Inoue2015}, $\sigma_t$ typically dominates over $c_s$. For the stellar component, we simply compute the radial velocity dispersion.

The last component required to calculate $Q_{gas}$ and $Q_{star}$ is the epicyclic frequency ($\kappa$) which is defined in terms of the local rotation velocity ($v_{\phi}$) and radius (R).

\begin{equation}
\kappa^2 = 2\frac{v_{\phi}}{R}\left(\frac{dv_{\phi}}{dR}+\frac{v_{\phi}}{R}\right)
\end{equation}

\cite{Inoue2015} discuss two possible approaches {for measuring $\kappa$}: separately for gas and stellar components using local rotation velocities, or combined for both components using a circular velocity profile ($v_{\phi} \approx v_{circ} = \sqrt{GM(<r)/r}$). We choose the latter method, as it ensures that $\kappa$ is never imaginary, therefore allowing us to automate the process of computing Q values for our large ($\sim10^5$) sample of subhaloes.

To combine $Q_{gas}$ and $Q_{star}$, several methods have been proposed. We follow the method of \cite{Romeo2011}. Specifically, for a combined $Q_{2comp}$:

\begin{equation}
\frac{1}{Q_{2comp}}=\begin{cases}
                     \frac{W}{Q_{star}}+\frac{1}{Q_{gas}},& (Q_{star}>Q_{gas})\\
                     \frac{1}{Q_{star}}+\frac{W}{Q_{gas}},& (Q_{star}<Q_{gas})
                    \end{cases}
\end{equation}

where

\begin{equation}
W=\frac{2\sigma_{star}\sigma_{gas}}{\sigma^2_{star}+\sigma^2_{gas}}
\end{equation}

\begin{figure}
\begin{center}
\includegraphics[width=3in]{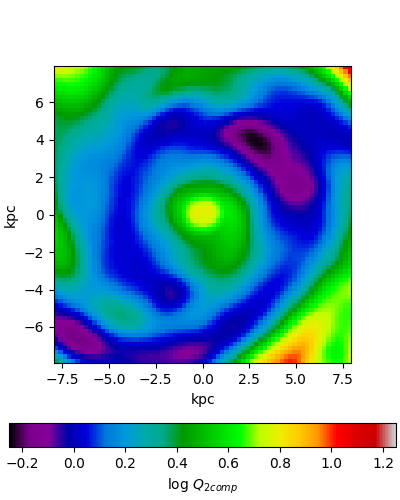}
\end{center}
\caption{Map of the {combined two-component} Toomre $Q_{2comp}$ for an example subhalo (Subfind ID 577184, $z=0$, $\Ms = 10^{10.0} \Msun$, sSFR $= 10^{9.6}$yr$^{-1}$). {Areas of low Q (purple/black) are Toomre unstable.}} 
\label{toomrecomb}
\end{figure}

\fig{toomrecomb} shows {a map of $Q_{2comp}$ for an example subhalo.} Finally, for quantitative comparison between subhaloes, we reduce each $Q$ map to a single number by computing the {{baryonic (stellar plus gas)}} mass fraction within 2$R_e$ where $Q_{2comp} < 2$.  Using $Q < 1$ instead of $Q < 2$ makes no qualitative differences to our results, however we chose $Q < 2$ to include all possible regions of instability \citep{Elmegreen2011}. 

\section{Results}
\label{results}

Our goal is to investigate galaxy structural evolution as measured by the relation of stellar mass ($\Ms$) to core stellar density ($\sigtwo$). We proceed in three steps. First, we check that TNG {{qualitatively}} reproduces the observational trends from WE19. Second, we track galaxies back in time and quantify the strength of their core growth by the slope of their evolution in the log $\sigtwo$-log $\Ms$ plane. Third, we investigate possible drivers of structural evolution by correlating various galaxy properties with past and future core growth. 

\subsection{Comparison {between TNG and Woo \& Ellison 2019}}
\label{compobs}

{Our first goal is to check whether TNG reproduces the observed trends under investigation, particularly those presented in WE19.  Specifically, we reproduce the $\sigtwo$-$\Ms$ relation and examine how other galaxy properties (AGN fraction, and the gradients of stellar age, sSFR, and gas metallicity) vary as a function of position in the relation.  {Note that WE19 studied $\sigone$, while we have preferred $\sigtwo$ for resolution reasons.  Therefore, wherever we present a comparison between TNG and the results of WE19, we have replotted their data using $\sigtwo$, leaving all other aspects of their analysis completely unchanged (qualitatively, the same observational trends are present regardless of whether $\sigone$ or $\sigtwo$ are used.) {{Note that although the gradients in the MaNGA sample were not PSF-corrected, they were selected to be nearby galaxies for which the PSF FWHM corresponds to less than 2 kpc in physical size, which is similar to the gravitational softening length in TNG of $\sim$1 kpc.}} For more details about the observational analysis, please see WE19.} 
} 

\begin{figure}
\includegraphics[width=0.9\linewidth]{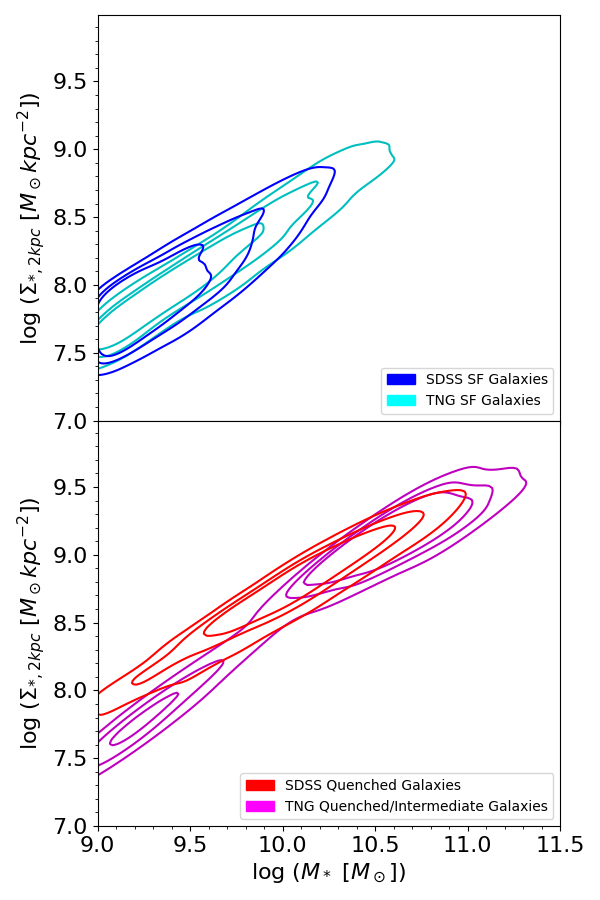}
\caption{\small The distribution of isolated galaxies in {central stellar mass} density ($\sigtwo$) vs {total} stellar mass ($\Ms$) in the SDSS and TNG, split into SF (top) and quenched/intermediate galaxies (bottom). Contour lines represent the same normalized number densities of galaxies for both SDSS and TNG, and are drawn at {{PDF = 0.1, 0.3, and 0.5}}. SDSS galaxy densities are corrected for luminosity.   
{{SF and quiescent}} galaxies in TNG are concentrated at a slightly higher $\sigtwo$ and $\Ms$, however they generally lie on the observed $\sigtwo$-$\Ms$ relation.}
\label{qtsfcent}
\end{figure}

{
First we {examine} the $\sigtwo$-$\Ms$ relation for TNG subhaloes, and isolated Sloan Digital Sky Survey (SDSS) galaxies in \fig{qtsfcent}.
}
The SDSS sample is identical to the sample used in WE19. TNG's distributions of {{SF and}} quiescent galaxies are skewed slightly to higher $\sigtwo$ and $\Ms$ than SDSS. {Otherwise, TNG subhaloes fall onto the observed relations {{fairly}} well.  We believe the good agreement is the result of at least the following reasons: 1) among several wind models explored for TNG, the fiducial model was chosen to reproduce the observed galaxy stellar size-mass relation at $z=0$ \citep{Pillepich2018}; 2) galaxy sizes and $\sigtwo$ are closely related for SF galaxies \citep{Chen2020}; and 3) the spatial resolution of TNG is roughly the same as that of SDSS, which has a diluting effect on $\sigtwo$}. {{We also note that TNG produces a distinct population of low mass quiescent subhaloes which appear to have quenched without kinetic mode BH feedback. While interesting, we defer investigation of these low-mass quenched galaxies for future work.}}

\begin{figure}
\includegraphics[width=0.9\linewidth]{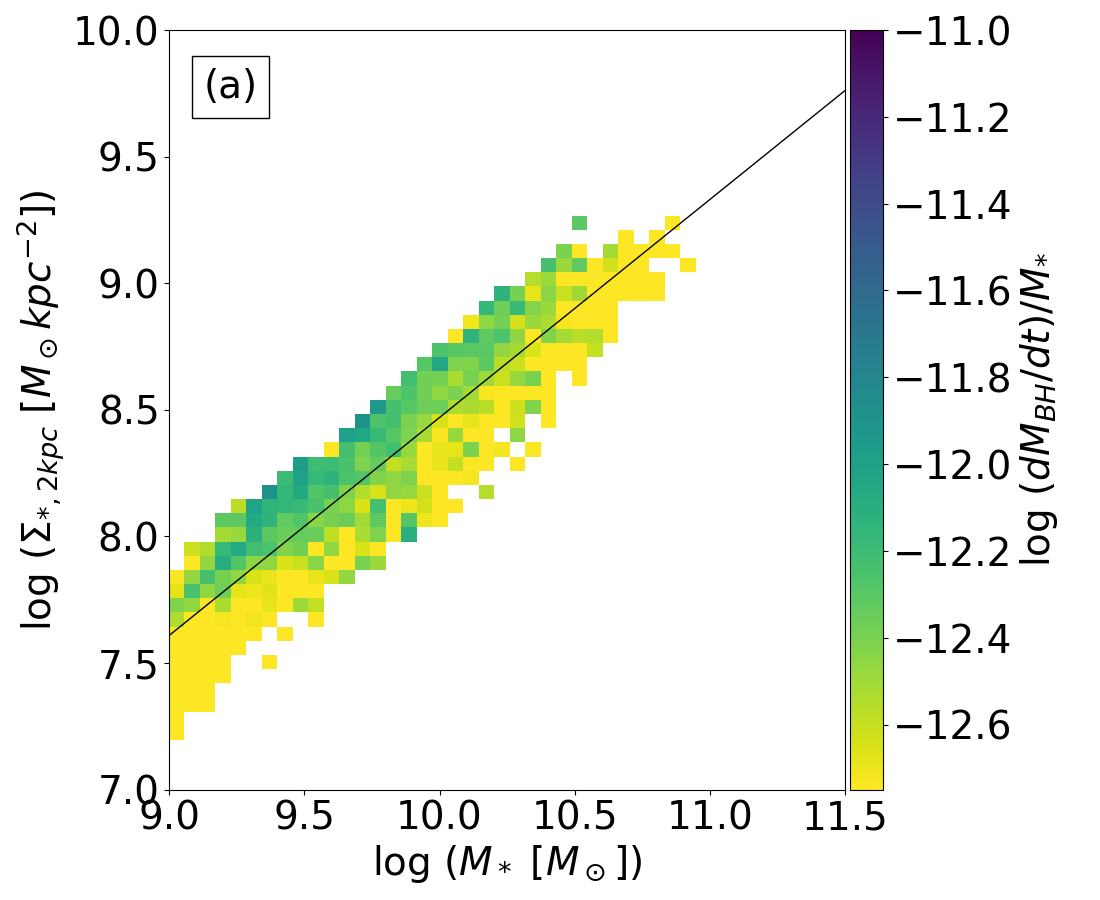}
\includegraphics[width=0.9\linewidth]{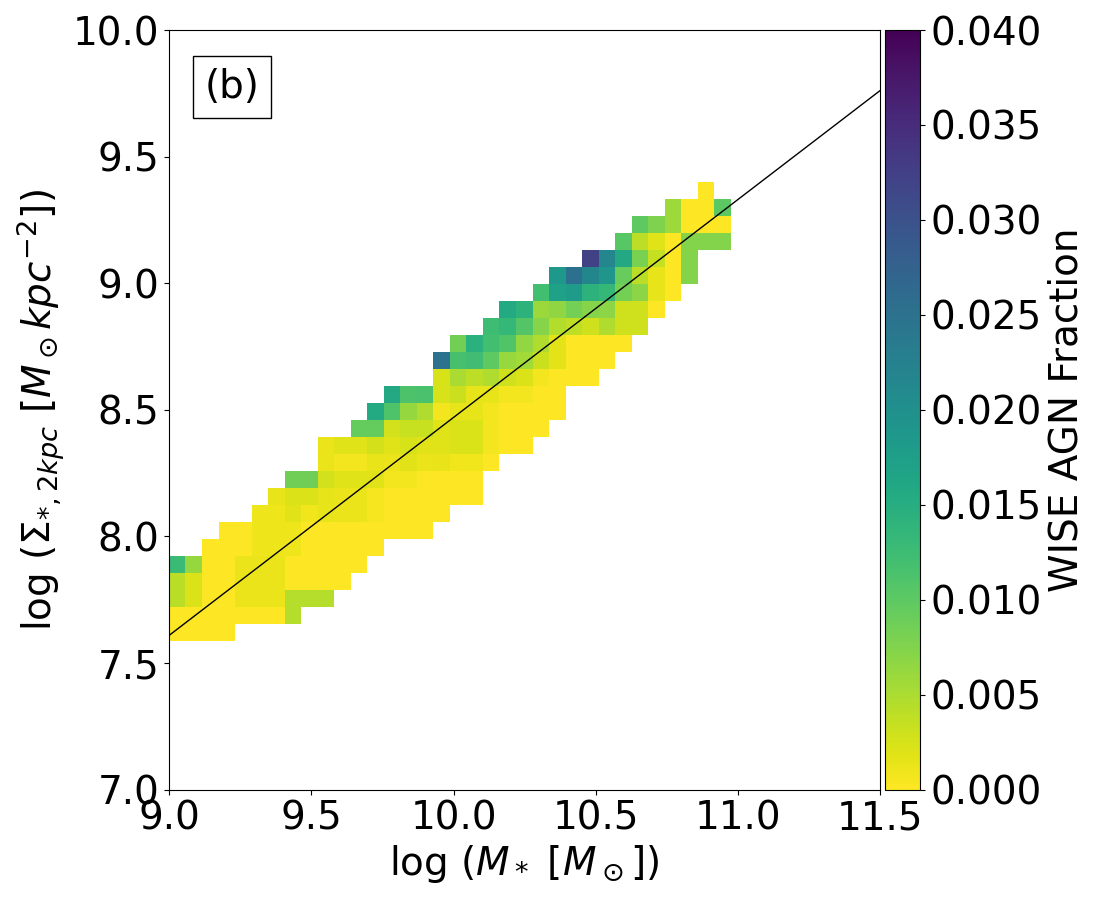}
\caption{The median black hole accretion rate normalized by the stellar mass {in TNG} (a) and {the WISE-detected AGN fraction for {{SDSS}} SF galaxies (b). Panel (b) is produced following the method of WE19 Fig. 10, but using $\sigtwo$ instead of $\sigone$}. The line shown, for comparison, is {the least squares fit to the TNG SF subhaloes}. {TNG shows good agreement, with higher AGN activity at higher $\sigtwo$ for a given $\Ms$.}}
\label{agncompare}
\end{figure}

\begin{figure}
\includegraphics[width=0.9\linewidth]{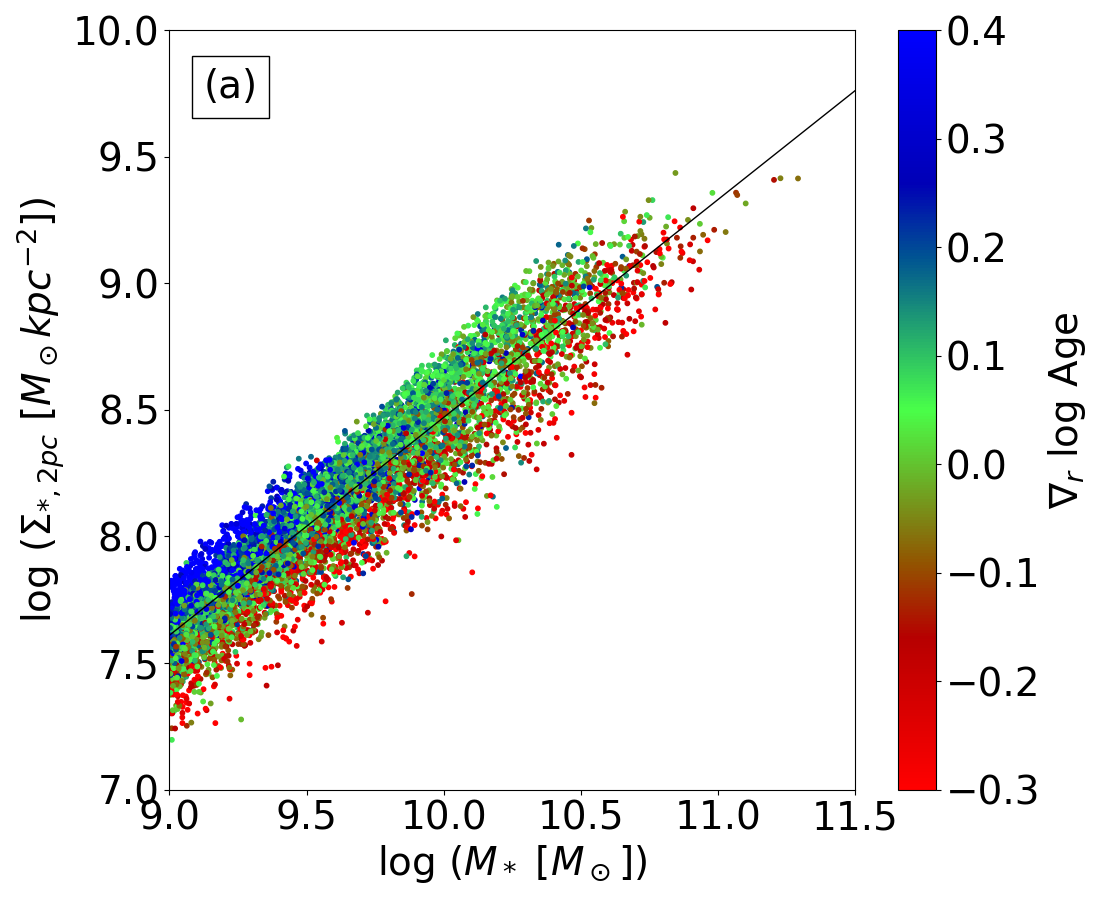}
\includegraphics[width=0.9\linewidth]{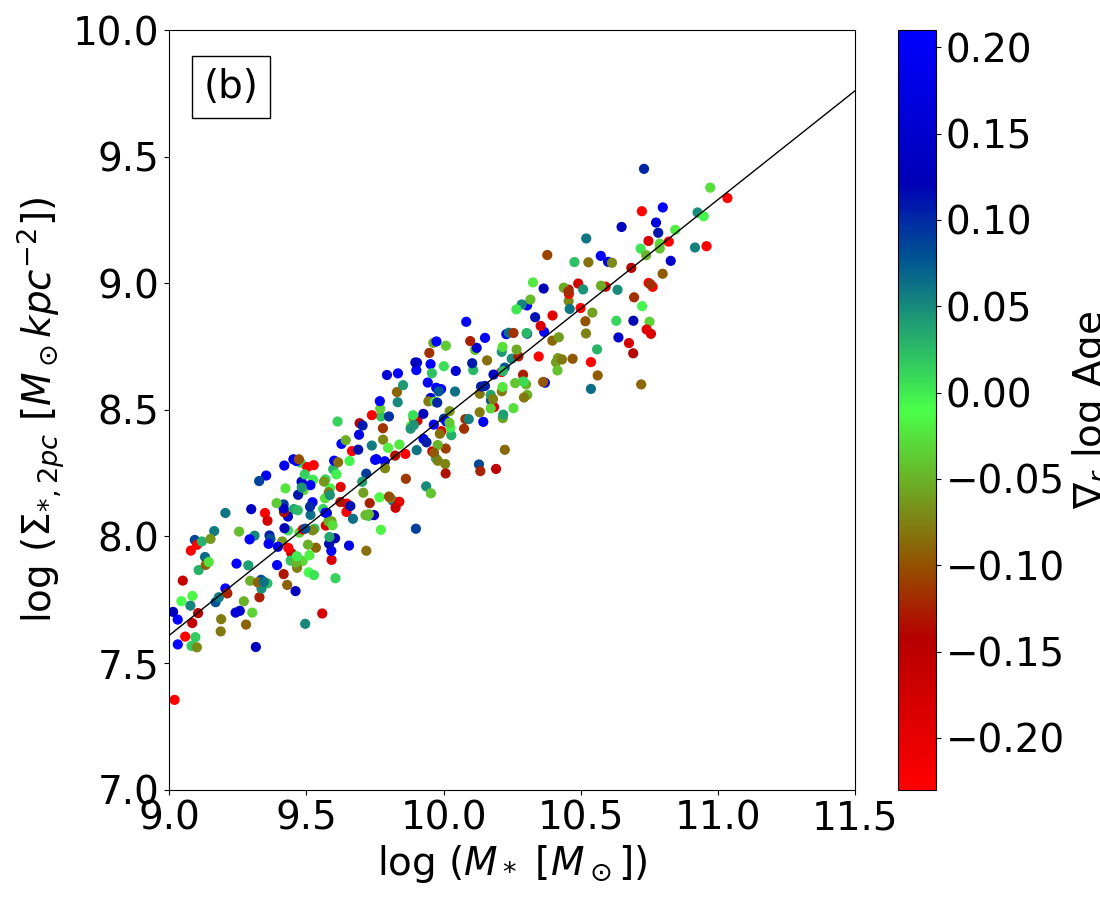}
\caption{Radial gradients of stellar age in TNG {(a) and MaNGA (b)} as a function of position in the core density {$\sigtwo$} vs stellar mass $\Ms$ plane. {Panel (b) is produced following the method of WE19 Fig. 3a, but using $\sigtwo$ instead of $\sigone$. The line shown, for comparison, is the least squares fit to the TNG SF subhaloes.} {TNG subhaloes} with higher {$\sigtwo$} at a fixed $\Ms$ tend to have younger cores compared to their outskirts, in qualitative agreement with findings from MaNGA. }
\label{radgrads1}
\end{figure}

\begin{figure}
\includegraphics[width=0.9\linewidth]{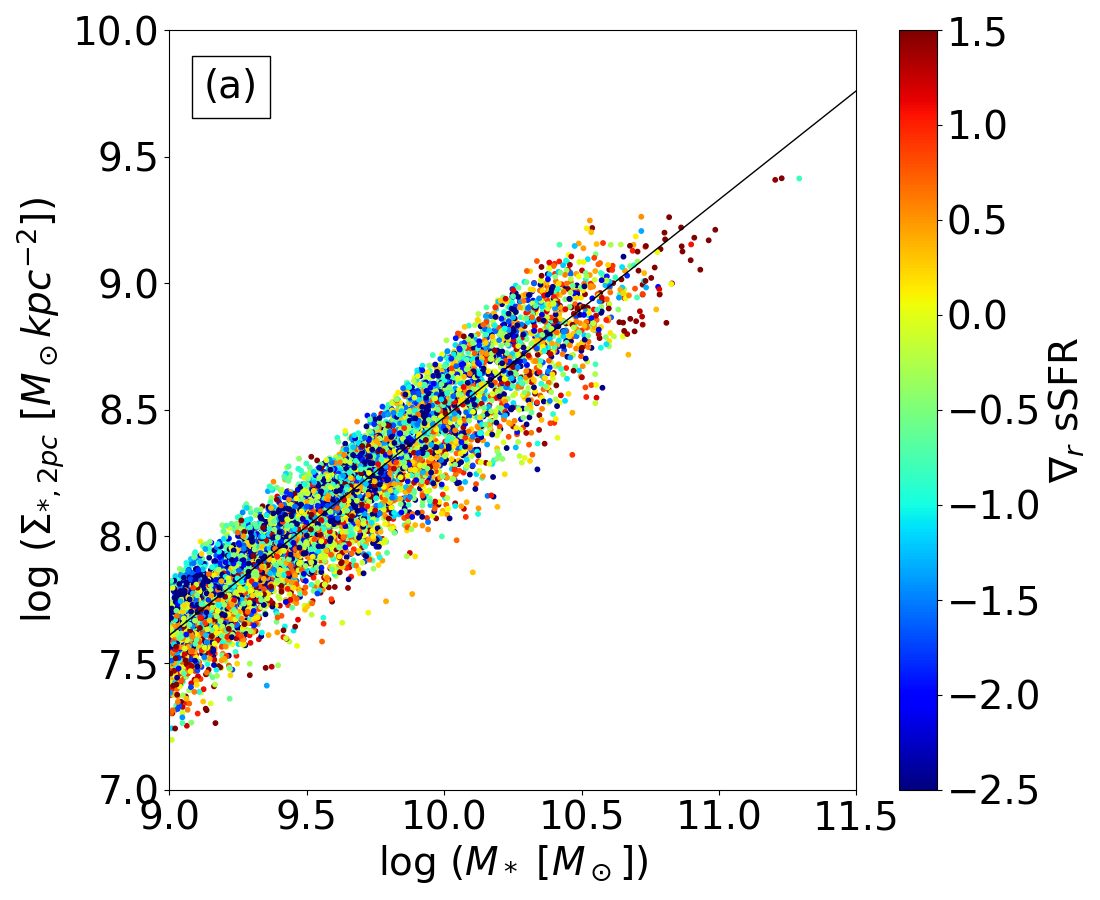}
\includegraphics[width=0.9\linewidth]{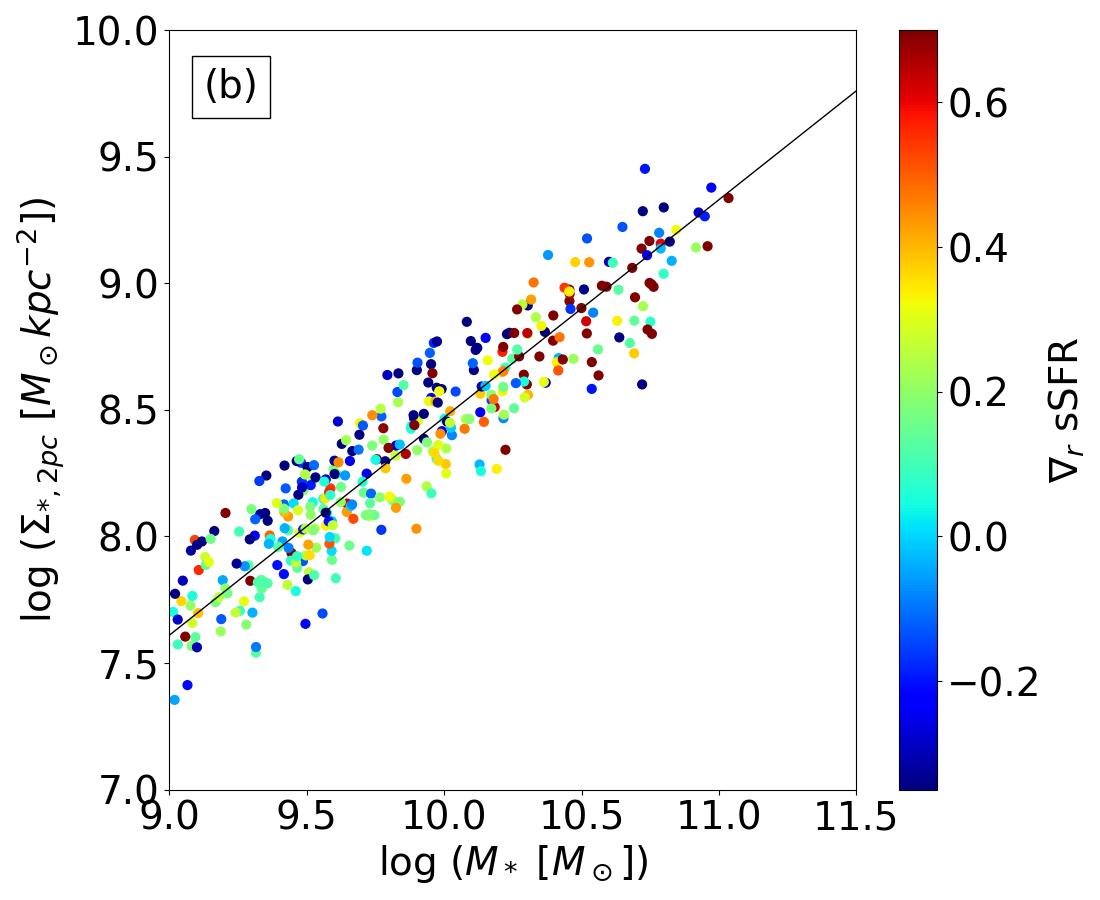}
\caption{Radial gradients of sSFR in TNG {(a) and MaNGA (b)} as a function of position in the central stellar density {$\sigtwo$} vs stellar mass $\Ms$ plane. {Panel (b) is produced following the method of WE19 Fig. 4a, but using $\sigtwo$ instead of $\sigone$. The line shown, for comparison, is the least squares fit to the TNG SF subhaloes.} {TNG subhaloes} with higher {$\sigtwo$} at a fixed $\Ms$ tend to have centrally concentrated sSFR, in qualitative agreement with findings from MaNGA.}
\label{radgrads2}
\end{figure}

\begin{figure}
\includegraphics[width=0.9\linewidth]{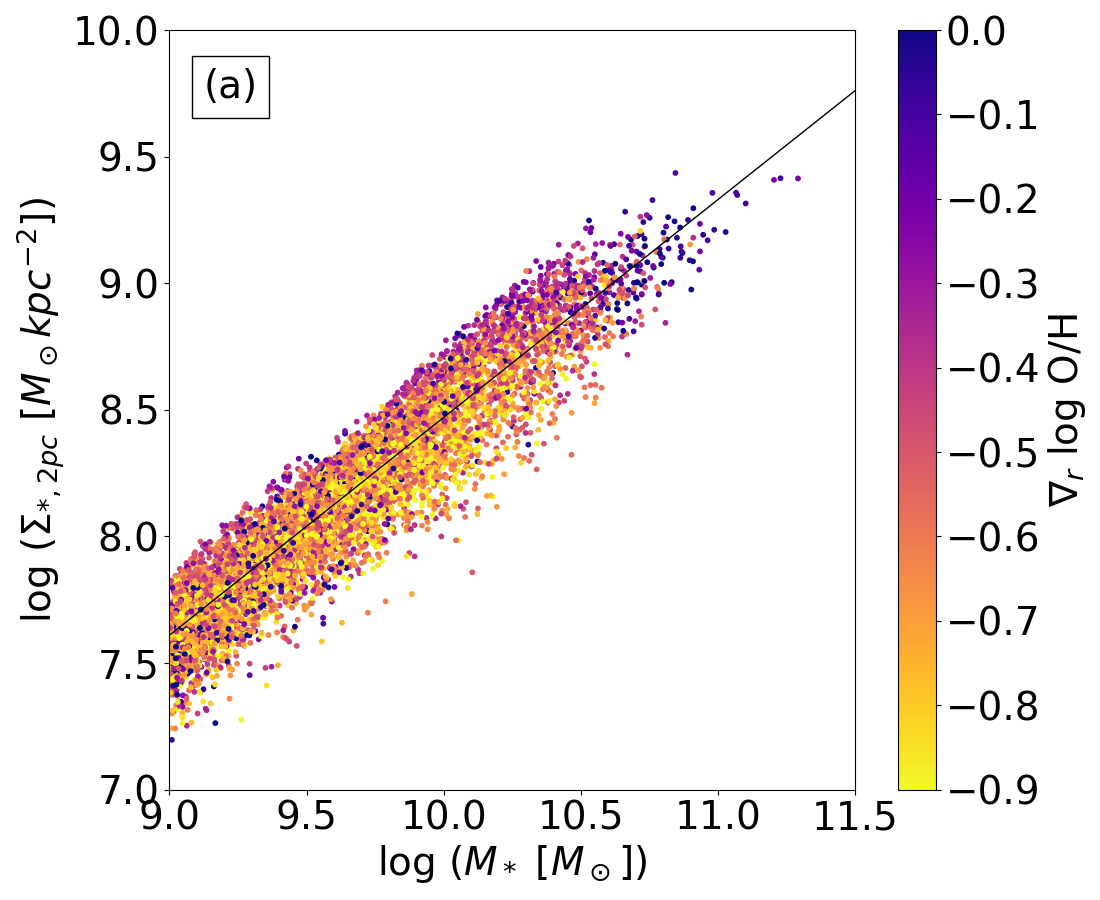}
\includegraphics[width=0.9\linewidth]{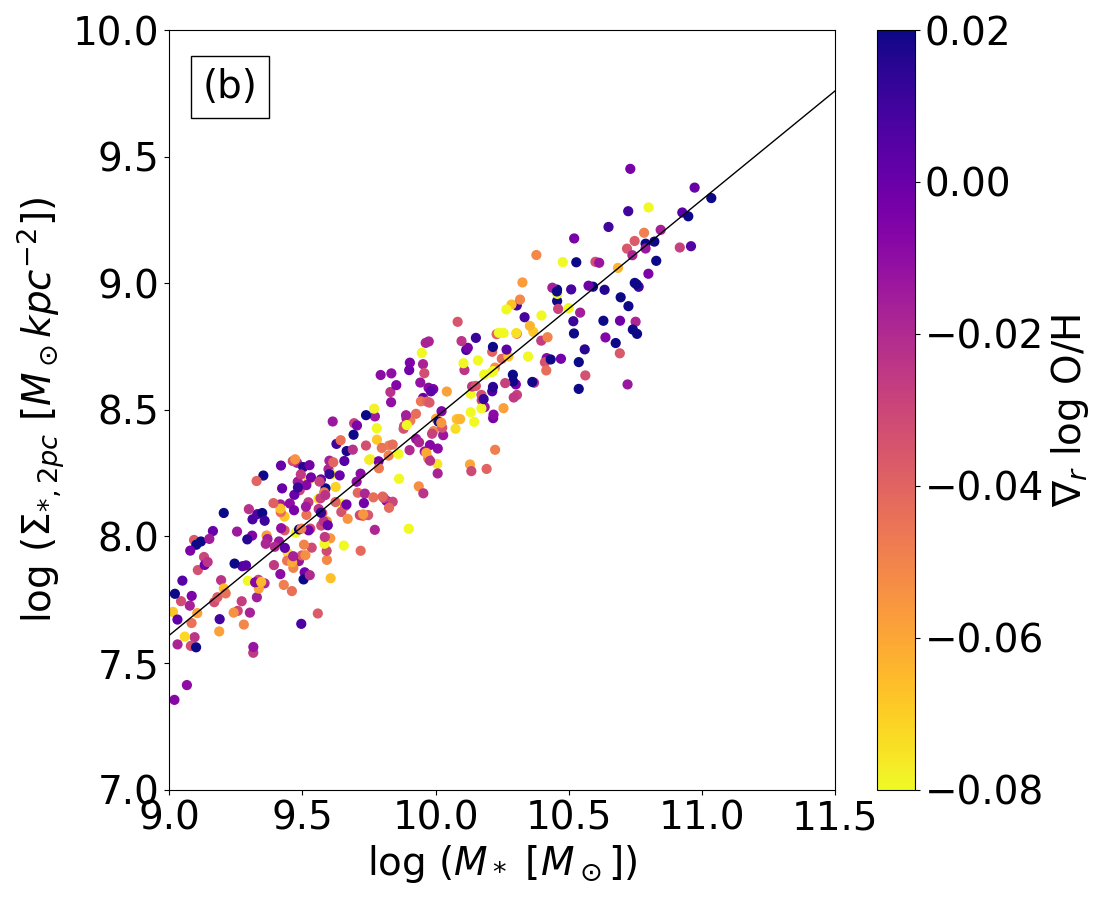}
\caption{Radial gradients of gas metallicity in TNG {(a) and MaNGA (b)} as a function of position in the core density {$\sigtwo$} vs stellar mass $\Ms$ plane. {Panel (b) is produced following the method of WE19 Fig. 5a, but using $\sigtwo$ instead of $\sigone$. The line shown, for comparison, is the least squares fit to the TNG SF subhaloes.} {TNG subhaloes} with higher {$\sigtwo$} at a fixed $\Ms$ tend to have flatter gradients in gas metallicity, in rough agreement with findings from MaNGA.}
\label{radgrads3}
\end{figure}

Next, we examine the distribution of AGN activity in the $\sigtwo$-$\Ms$ plane {for SF galaxies only}. In WE19, the presence of an AGN was inferred using mid-IR colours from WISE. As a surrogate for AGN strength, we use the total BH accretion rate. When more than one BH particle is present in a subhalo, we sum all the accretion. Although the BH accretion rate does not tell us which mode the AGN is in (kinetic or thermal), the feedback in either case is directly proportional to it. \fig{agncompare} shows the BH accretion rate (panel a) and the BH mass (panel b), both normalized by $\Ms$, for isolated SF subhaloes, mapped to the $\sigtwo$-$\Ms$ plane. In all comparisons between observations and TNG (Figs \ref{agncompare}-\ref{radgrads3}), we draw a least-squares fit to the TNG SF subhaloes, using randomly projected values for $\sigtwo$: $\log(\sigtwo) = 0.86 (\log \Ms - 10.5) + 8.90$.  The accretion rate in TNG increases {with core density (at fixed $\Ms$), in agreement with the observed AGN fraction from WE19.} (Note that \fig{agncompare} uses the SDSS sample in WE19 rather than the MaNGA sample since spatial information was not needed.)

As a final comparison between simulation and observational trends, we compare radial gradients of stellar age {(\fig{radgrads1})}, sSFR {(\fig{radgrads2})}, and gas metallicity {(\fig{radgrads3})} in TNG to the results of WE19 using their MaNGA sample. We note that, due to the different methods used to calculate radial gradients, we only compare our results to WE19 qualitatively. The simulated gradients of age and sSFR are in qualitative agreement with the observations. Galaxies with higher core densities (at fixed $\Ms$) have younger stars, and have elevated sSFR in their cores. {TNG produces a weak radial trend in metallicity gradient, with flatter gradients at higher $\sigtwo$ in agreement with WE19.} 

\subsection{Galaxy History in the $\sigtwo$-$\Ms$ plane}
\label{history}

Having demonstrated that TNG {generally reproduces} the observed patterns in the $\sigtwo$-$\Ms$ plane, we next examine the evolution of subhaloes over time {to determine why the age and sSFR gradients correlate with $\sigtwo$.  
WE19 hypothesized that galaxies with high $\sigone$ are undergoing a compaction-like process which brings gas towards the centre of the galaxy, increasing the central sSFR, lowering the mean stellar age, and increasing $\sigone$.  In contrast, galaxies with shallow (or positive) sSFR gradients do not build $\sigtwo$ as quickly as their total mass.  As a result, the evolutionary paths of galaxies in the $\sigtwo$-$\Ms$ diagram are expected to be shallower for those galaxies with low $\sigtwo$, and steeper for those with higher $\sigtwo$.  These expectations can be tested directly in TNG.  

{In the current work, we study} galaxy evolution in the $\sigtwo$-$\Ms$ diagram} 
 in two ways. First, we examine the evolution of subhaloes as a population. Second, we {track} the evolution of individual subhaloes. \fig{subanim} shows the evolution of TNG subhaloes in the $\sigtwo$-$\Ms$ plane between $z = 2.0$ to $z = 0$. The selection (i.e. isolated SF subhaloes) is repeated at each snapshot.  The intercept of the distribution at $\Ms = 10^{10.5}$ decreases by $\sim$ 0.2 dex between $z=2.0$ and $z=0.5$ (10 and 5 Gyr ago), in agreement with results from observations {for $\sigone$} \citep{Barro2017,Chen2020}. The slope of the relation also decreases with decreasing redshift over this time (by $\sim$ 0.1, or 10\%). Although small, this flattening is statistically significant and disagrees with observations{, which show a roughly constant slope \citep{Barro2017,Chen2020}. In this case we are comparing $\sigtwo$ trends from TNG to $\sigone$ trends from observations. However, we did check $\sigone$ in TNG and obtain the same trends as with $\sigtwo$. {We discuss the implications of TNG's evolving slope of the $\sigtwo$-$\Ms$ relation in \secref{discussion}.}

\begin{figure*}
   \animategraphics[loop,autoplay,width=0.9\linewidth]{12.5}{figures/sigfitanim_files/sigfit_sig2_snap_}{33}{99} 
  \caption{\small Animation of the distribution of isolated subhaloes in core density $\sigtwo$ vs stellar mass $\Ms$ between $z=2$ and $z=0$. {The parameters of a least squares fit to the SF subhaloes, and the residual standard deviation are shown in the bottom right corner of the left panel. As subhaloes transition to quiescence they appear in the right hand panel.} {The slope of the fit decreases with decreasing redshift, in contrast to observations, which show a roughly constant slope. Otherwise, TNG {{qualitatively}} reproduces the extent and intercept of the SF and quiescent relations fairly well back to at least $z = 2.0.$} The animation embedded in this PDF file can be viewed in Adobe Reader.  }
\label{subanim}
\end{figure*}

We now shift focus to individual galaxies. As described in \secref{trackhist} we trace the histories of $z=0$ galaxies up to $z=0.5$. \fig{hists} shows the evolutionary histories of 9308 $z=0$ isolated SF subhaloes in the $\sigtwo$-$\Ms$ plane. The three {panels} represent three groups divided by their final ($z=0$) value of $\sigtwo$ relative to the {least squares fit in $\log\sigtwo - \log\Ms$ for all SF subhaloes, using face-on values for $\sigtwo$.} The line used is {$\log(\sigtwo) = 0.82 (\log \Ms - 10.5) + 8.80$.} {Note that this is slightly different than the fit reported in \secref{compobs}, which used randomly projected values for $\sigtwo$ instead of face-on values.} {We split our sample into subhaloes with dense (high $\sigtwo$, \fig{hists}(a)), average (\fig{hists}(b)), and diffuse (low $\sigtwo$, \fig{hists}(c)) cores for a given $\Ms$. } For visual clarity, each figure only shows a random sample of 100 subhaloes, {with each subhalo's history coloured randomly for ease of viewing}. 

{We make two observations from \fig{hists}. First, the slopes of the individual tracks are steepest in panel (a) and shallowest in panel (c). {\it Galaxies with dense cores have steeper evolutionary paths in $\sigtwo$-$\Ms$ than galaxies with diffuse cores.} This confirms WE19's interpretation of sSFR and stellar age gradients in MaNGA. Second, at low $\Ms$ ($\ltsima 10^{9.5} \Msun$), some galaxies follow much steeper trajectories indicating strong core growth relative to overall growth.}

\begin{figure}
\includegraphics[width=0.88\linewidth]{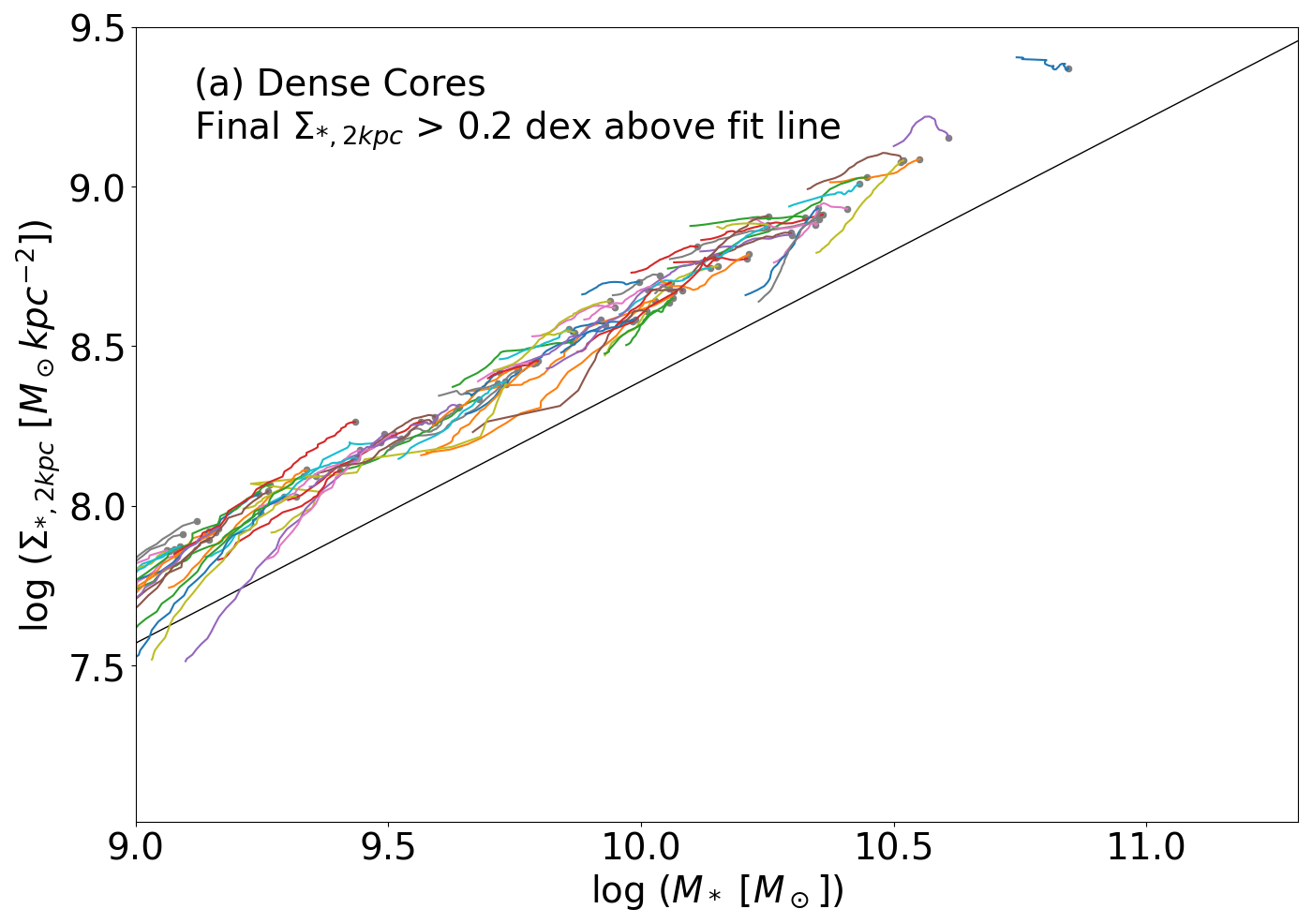}
\includegraphics[width=0.88\linewidth]{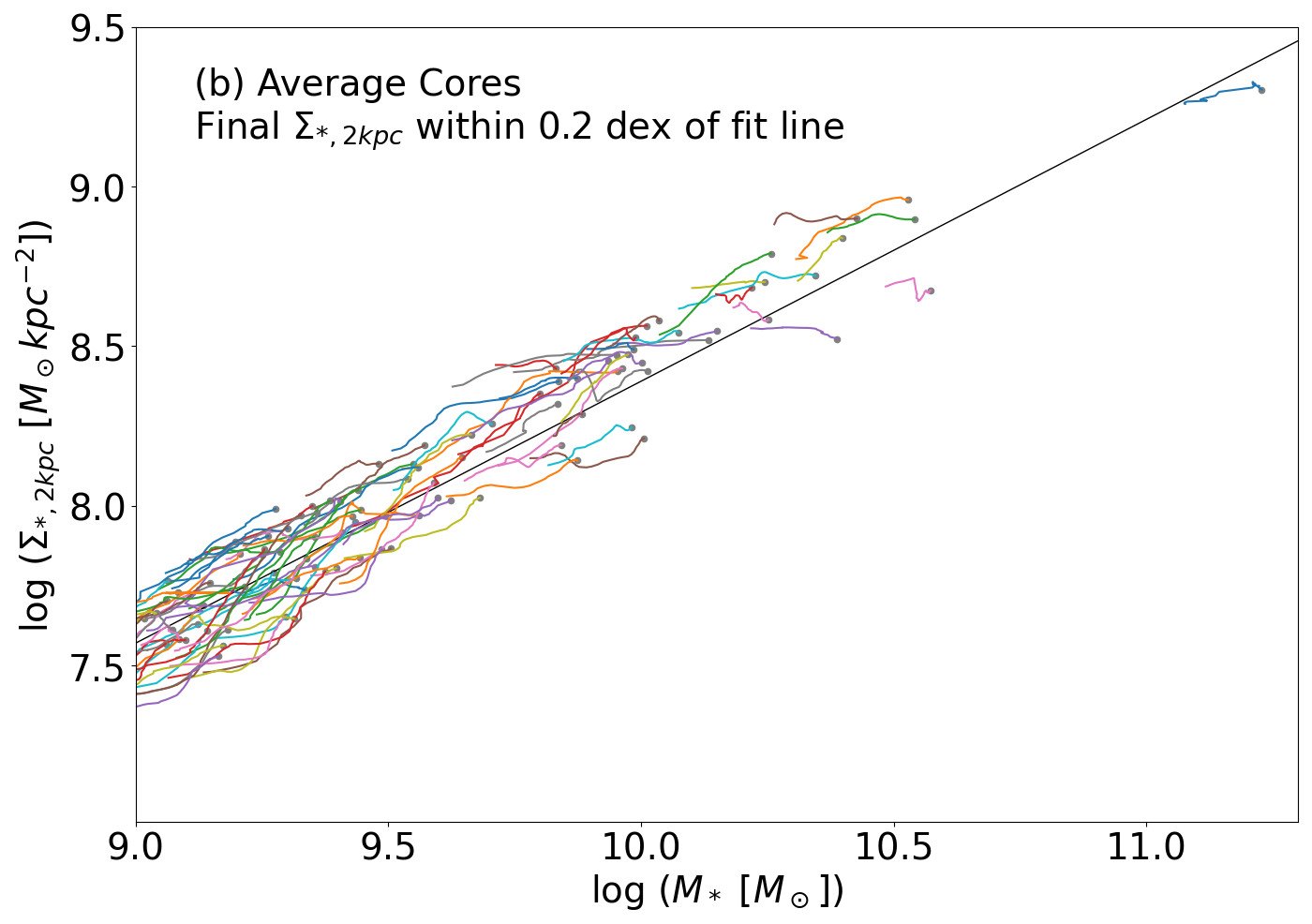}
\includegraphics[width=0.88\linewidth]{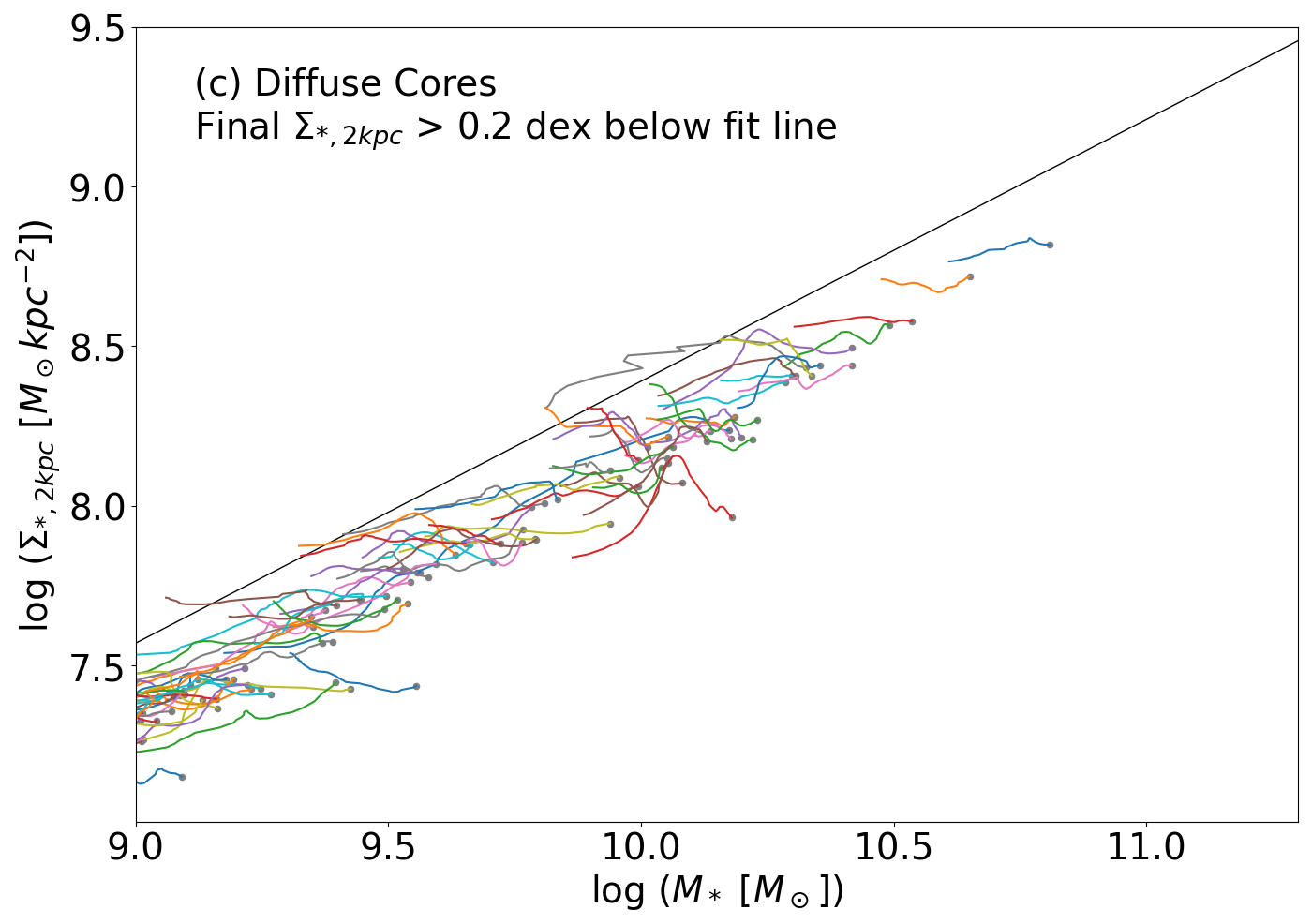}
\caption{The evolutionary trajectories of example TNG galaxies in the core density $\sigtwo$ vs stellar mass $\Ms$ plane.  The trajectories trace the histories of 100 subhaloes that end their evolution with high (panel a), intermediate (panel b), and low (panel c) $\sigtwo$ relative to their total $\Ms$. {Galaxies are randomly coloured for ease in distinguishing individual trajectories. Galaxies with denser cores (a) generally grow along steeper trajectories than those with diffuse cores (c). At low mass, some galaxies grow very steeply (panel a, $\Ms < 10^{9.5} \Msun$).} The line, for visual reference, is a least squares fit to all isolated SF subhaloes at $z=0$ (using face-on values for $\sigtwo$) and is identical in all panels. {{For visual clarity in this figure,}} we smooth the $\sigtwo$ values over time using a Gaussian kernel with a width of 2 snapshots.}
\label{hists}
\end{figure}

To quantify these observations, we measure the average slopes of the evolutionary tracks in the $\sigtwo$-$\Ms$ plane. The slopes are taken linearly between each subhalo's state at $z = 0.5$ and $z = 0$. These slopes correspond to the degree of compaction the subhalo underwent: i.e., subhaloes with steeper slopes grew their cores faster relative to their total stellar mass than those with shallower slopes. \fig{slopehist} shows two histograms of the slopes, split by total stellar mass at $\Ms = 10^{9.5} M_{\odot}$ to highlight the steeper evolutionary slopes possible at low $\Ms$ (\fig{hists} panels (a) and (b)). We further split each histogram based on $\sigtwo$ at $z = 0$, using the same three divisions used in \fig{hists}{: dense cores (blue), average cores (green), and diffuse cores (red).} Larger values of slope (x-axis) indicates stronger stellar core growth (compared with overall stellar mass growth).

\begin{figure}
\includegraphics[width=0.9\linewidth]{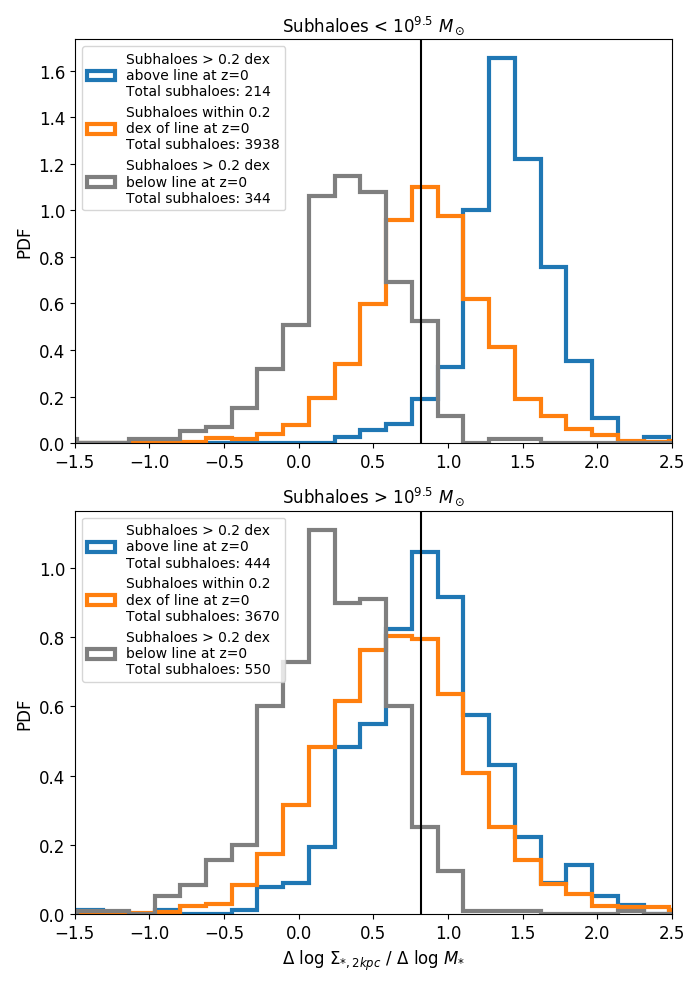}
\caption{Distributions of the average slope of galaxy trajectories on the core-density $\sigtwo$ vs stellar mass $\Ms$ plane. The panels show isolated star-forming subhaloes with $\Ms < 10^{9.5} M_{\odot}$ (top) and $\Ms > 10^{9.5} M_{\odot}$ (bottom). The trajectories track the main progenitor branch from $z = 0.5$ to $z = 0$.  The black vertical line at {{0.82}} marks the slope of the entire $\sigtwo$-$\Ms$ distribution at $z=0$. Numbers in the legend indicate the total number of subhaloes in each of the 6 groups.}
\label{slopehist} 
\end{figure}

\fig{slopehist} demonstrates the mass dependence of the evolutionary slopes, especially for galaxies with denser cores.  Galaxies less massive than $10^{9.5} M_{\odot}$ evolve more steeply in $\sigtwo$-$\Ms$ than more massive galaxies.  {We} suspect that the mass dependence of the core growth trajectories may be a resolution effect linked to weaker BH feedback in the less massive galaxies.  We confirm that in the lower resolution simulation TNG100-2, the steeper tracks extend to higher masses.

Second, \fig{slopehist} shows that at all masses, {\it galaxies with denser cores (blue histogram) {at $z = 0$} evolve along steeper trajectories in the $\sigtwo$-$\Ms$ plane compared to galaxies with diffuse cores (grey histogram)}. Third, most ($\sim$65\%) subhaloes with $\Ms > 10^{9.5} M_{\odot}$ evolve along slopes that are shallower than the slope of the single-snapshot distribution {fit line} of 0.82. In fact, the overall median slope between $z$ = 0.5 and $z$ = 0 of isolated SF subhaloes with $\Ms >10^{9.5} M_{\odot}$ is {only} 0.63. {In other words, the median galaxy does not evolve parallel to the overall distribution at any single snapshot, but rather follows a more shallow trajectory.}

{We note that a small proportion of galaxies have slightly negative slopes. In these cases, the density of stars in their inner 2 kpc actually decreased as they grew in overall $\Ms$. This could be due to stellar evolution and death, numerical effects (integration errors over time, misplaced subhalo centre points after disruptions, etc), or also some adiabatic expansion {(see e.g. \citealp{Wellons2016})}.} 

\subsection{The Drivers of Structural Evolution}

Having shown that galaxies with different evolutionary histories preferentially populate unique regimes in the {$\sigtwo$}-$\Ms$ plane, we next investigate the possible drivers of these distinct evolutionary trajectories. {{Our objective is to test why some subhaloes have an excess of gas that leads to core growth.}} In particular, we investigate the {role} of {the sSFR gradients,} mergers, disc instabilities, black hole feedback and the angular momentum of accretion on galaxy core growth.  We quantify the core growth using the average slope of the evolutionary track on the $\sigtwo$-$\Ms$ diagram over the last 2 Gyr or 4 Gyr and correlate this slope with the desired property at the start and end points of the track.  
{To increase our sample size and achieve more robust statistical results, we consider ``$z=0$'' to include subhaloes from snapshots up to $z = 0.05$ (5 snapshots total). {{In other words, we take 5 different evolutionary segments from each galaxy and treat them as distinct data points. This produces consistent results to analyzing only the last snapshot of subhaloes, but with smoother trends.}}}

The effects of galaxy mergers are treated slightly differently as detailed in Section \ref{mergers}.

\subsubsection{{The sSFR Gradient and Core Buildup}}
\label{correlate}

The observation that galaxies with dense cores have centrally elevated sSFR led WE19 to suggest that central star-formation increases the density of the core and leads to steep evolutionary paths in the $\sigtwo$-$\Ms$ diagram. {Indeed, we expect most drivers of core buildup to create a negative sSFR gradient, leading to increased $\sigtwo$. The only other possibility would be inwards migration of stars, which we expect to be rare.} {In other words, we expect that the sSFR gradient is the {\it mechanism} by which other fundamental drivers determine galaxy evolution in $\sigtwo$-$\Ms$.} 

We test this explicitly in \fig{slopesfrslope} (left) where we show the average slope of the evolutionary path in $\sigtwo$-$\Ms$ as a function of the sSFR gradient at the start (blue) and end (orange) of the 2-Gyr evolutionary track.  The right panel shows the same, but over a time interval of 4 Gyr.  The blue curves of \fig{slopesfrslope} shows that at the start of the evolutionary tracks, the sSFR gradient {correlates weakly} with the subsequent growth of the core.  Galaxies with positive sSFR gradients $\sim 2$ Gyr ago (SF occurs in the outskirts) were {somewhat} more likely to grow their total mass more than their cores (shallower evolutionary paths).  {However right panel shows that correlation between core growth and sSFR gradient weakens the farther back we measure the sSFR gradient, virtually disappearing for sSFR gradients measured 4 Gyr ago.} 

{If we look instead at the sSFR gradients at the {\it end} of the evolutionary tracks (the orange curves), the sSFR gradient correlates more strongly with the past growth of the core.  This correlation between the end-point sSFR gradient and core growth also weakens for longer time intervals, but does not disappear after 4 Gyr.  In other words, {\it the current sSFR gradient is an indicator of past core evolution} (over at least 4 Gyr), but past sSFR gradient is not a major predictor of future core growth.} 

\begin{figure*}
\includegraphics[width=0.45\linewidth]{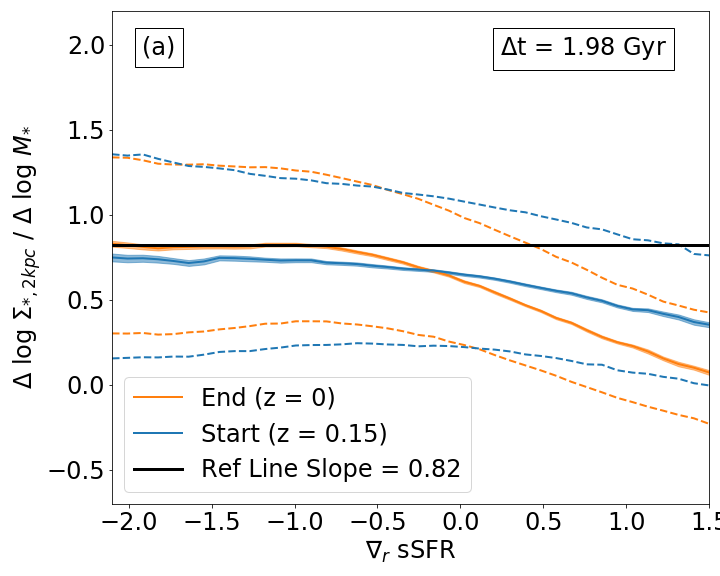}
\includegraphics[width=0.45\linewidth]{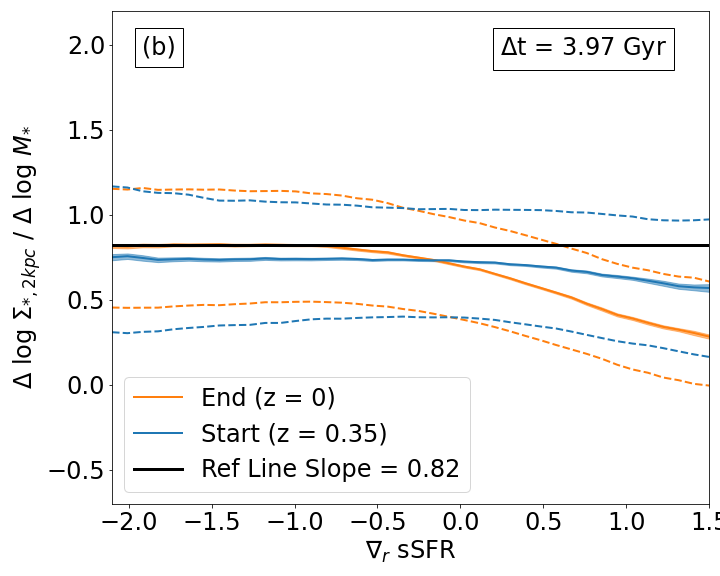} 
\caption{The running median values of the slope of the evolutionary trajectory in the $\sigtwo$-$\Ms$ plane as a function of galaxy sSFR gradient at the start (blue) and end (orange) of a 1.98 Gyr (left) and 3.97 Gyr (right) trajectory.  Quartiles are indicated as dashed lines.
The black reference line shows the slope of the {least-squares fit to the} $\sigtwo$-$\Ms$ relation at $z=0$.  The sSFR gradient at $z=0$ is an indicator of {\it previous} core growth over at least 4 Gyr.
However the sSFR gradient is not a strong predictor of future core growth, with any predictive power disappearing after 4 Gyr.}  
\label{slopesfrslope}
\end{figure*}

\subsubsection{Galaxy Mergers}
\label{mergers}

Galaxy mergers are a strong candidate for driving core growth. Simulations predict that major mergers can trigger central starbursts {(e.g. \citealp{Cox2008}, \citealp{Moreno2015})}. However, because we select isolated subhaloes at $z = 0$, only a small fraction of our sample ($\sim$ 10\%) have had major mergers since $z = 0.5$. For those subhaloes that did undergo major mergers during this period, we examine the effect the merger had on their evolution in the $\sigtwo$-$\Ms$ plane. We quantify the core growth (by slope in $\sigtwo$-$\Ms$) in a time window $\Delta t_{pre/post}$ before and after the merger. We then compare the two slopes.

\Fig{mergereffect} shows 3 histograms of the difference in trajectory slopes before and after a merger, where we use a $\Delta t_{pre/post}$ of 4 snapshots ($\sim$ 650 Myr) in 3 mass bins (we vary $\Delta t_{pre/post}$ between 2 and 6 snapshots and find no major variation in our basic results). {{We divide our sample of mergers in half based on the gas fraction ($f_{gas}$) within 2$R_e$ of the second most massive progenitor immediately prior to the merger (bottom two frames of \Fig{mergereffect}). The median value of $f_{gas}$ for our sample is 0.40. We also tried dividing based on the total gas fraction of the first two most massive progenitors, as well as using the total bound gas fraction, and found no substantive difference. Mergers are generally wetter at lower $\Ms$. Based on our criteria, 68\% of low-mass, 51\% of mid-mass, and only 20\% of high-mass mergers are ``wet.''}}

{\fig{mergereffect} shows that whether or not mergers are likely to induce rapid core growth is mass- {{and $f_{gas}$-dependent. Wet mergers, on average, lead to increased core growth at all masses. Lower mass galaxies experience a stronger effect. Dry mergers have less effect, with high mass ($\Ms \gtsima 10^{10} \Msun$) dry mergers actually resulting in shallower evolutionary trajectories.}}
However, since mergers are rare overall, they are unlikely to be the main driver of rapid core growth in the isolated galaxy population.

\begin{figure}
\includegraphics[width=0.95\linewidth]{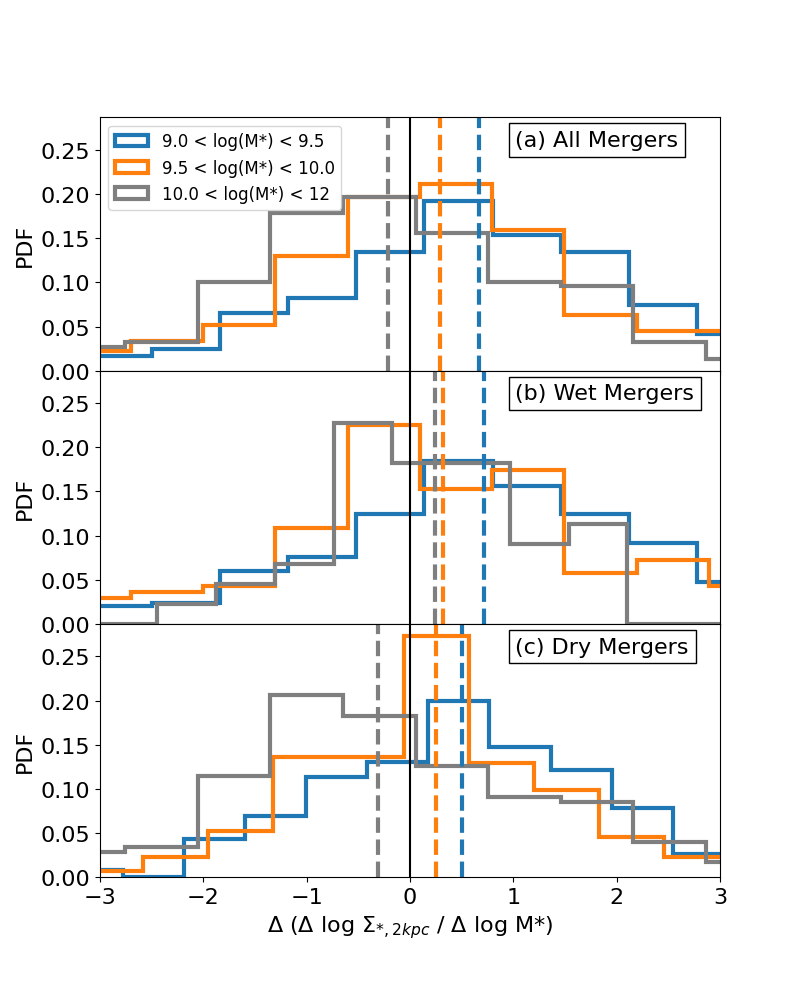}
\caption{The distribution of trajectory changes after a major merger,  {where a value of zero (black vertical line) indicates no change. Positive values indicate stronger core growth, whereas negative values indicate weaker core growth, relative to total $\Ms$ growth.} {{Vertical dashed lines show median values.}} 
Histograms include only isolated star-forming subhaloes that underwent major mergers since $z = 0.5$. {{We divide our sample of mergers in half based on the gas fraction ($f_{gas}$) of the second most massive progenitor: (a) shows all mergers, (b) shows mergers with $f_{gas} > 0.40$, and (c) shows mergers with $f_{gas} < 0.40$. Mergers, on average, lead to increased core growth, except for high mass dry mergers which have slightly the opposite effect. Core growth tends to be slightly stronger at lower mass and higher $f_{gas}$. }}}
\label{mergereffect}
\end{figure}

\subsubsection{Disc Instability and Core Buildup}
\label{toomredisc}

Another candidate for driving the structural evolution of isolated galaxies is the instability of the disc components (both stellar and gas). Disc instabilities lead to fragmentation into clumps of star-formation, and the inflow of star-forming gas towards the galaxy centre, resulting in the build-up of the core {{\citep{Dekel2014a,Zolotov2015,Inoue2015}.}}
 
To investigate disc instability as a driver of core build-up, we construct maps of the Toomre $Q$ parameter (on a resolution scale of 0.74 kpc) for each subhalo and its main progenitors. We then compute the fraction of mass within 2$\Re$ that is Toomre unstable, i.e., with $Q < 2$ (as discussed in detail in {Section} \ref{toomre}).    
We also try using $R_e$ instead of 2$R_e$ and using a cutoff $Q$ of 1 instead of 2, and in all cases obtain similar results.

\fig{toomrecorrelate} shows the correlation of evolutionary slopes in $\sigtwo$-$\Ms$ with the unstable mass fraction at the start (blue) and at the end (orange) of evolution segments of 2 Gyr (the results are similar for 4 Gyr). The blue curve shows that the unstable mass fraction at the start of the evolution has almost no predictive power on the steepness of the core growth in $\sigtwo$-$\Ms$. In contrast, the orange curve shows that the unstable mass fraction at the end of the evolution correlates negatively with the steepness of core growth.  In other words, galaxies that built their cores rapidly over the last 2 Gyr end up with more stable discs. This can be understood as the stabilising effect of central spheroids as pointed out by \cite{Martig2009,Gensior2020}.  Therefore, {\it disc instabilities are not a strong driver of structural evolution in TNG}. Instead, it is structural evolution that affects the stability of the disc.

\begin{figure}
\includegraphics[width=0.9\linewidth]{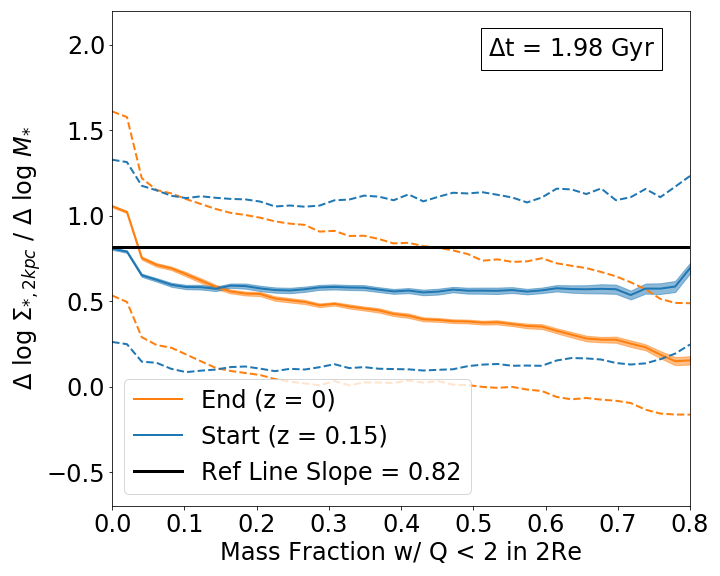}
\caption{The running median values of the slope of the evolutionary trajectory in the $\sigtwo$-$\Ms$ plane over the last 1.98 Gyr as a function of the unstable mass fraction at the start (blue) and end (orange) of the trajectory.  Quartiles are indicated by the dashed lines. The black reference line shows the slope of the {least-squares fit to the} $\sigtwo$-$\Ms$ relation at $z=0$.
 Disc instability $\sim 2$ Gyr ago has no significant predictive power for core growth.  However more rapid core growth over the last 2 Gyr predicts more stable discs today.}
\label{toomrecorrelate}
\end{figure}

\subsubsection{Black Hole Feedback and Core Buildup}
\label{bhdisc}

Supermassive black hole feedback is another candidate for influencing the structural evolution of isolated galaxies.  We expect the growth of the BH to {potentially} have {two opposing} effects on $\sigtwo$.  On the one hand, the growth the BH should correlate with the growth of $\sigtwo$ since the gas inflows that build the latter also feed the former \citep{Fang2013,Chen2020}. On the other hand, in TNG, BH feedback returns energy and/or momentum to the galactic gas resulting in a suppression of star formation \citep{Weinberger2018,Pillepich2018,Nelson2019,Terrazas2020,Zinger2020}.  Two modes of feedback are implemented in TNG corresponding to a high-accretion or thermal mode that injects energy into the gas cells in the feedback region, and a low-accretion or kinetic mode that injects a momentum boost to the gas (see \citealp{Weinberger2018} and \citealp{Zinger2020} for details).  The feedback region is 2-3 kpc in size, so even if the feedback does not succeed in quenching the entire galaxy (quenching seems primarily due to the kinetic mode - \citealp{Weinberger2018,Terrazas2020}), BH feedback plausibly suppresses core growth, even if star formation continues outside the feedback region.   

We note with caution that in TNG, BH masses for star-forming galaxies are too massive compared to observed BH masses by almost 2 orders of magnitude {(see the top left panel of Fig. 7 in \citealp{Terrazas2020})}.  For example, TNG predicts that a star-forming galaxy comparable in mass to the Milky Way should have a supermassive BH of mass $\gtsima 10^{8.5}\Msun$ compared to the Milky Way's BH of $\sim 10^{6.6}\Msun$ \citep{Ghez2008,Genzel2010}.  So although the $\MBH$-$\Ms$ relation is reproduced for the quiescent population \citep{Pillepich2018}, BHs seem to grow too quickly in star-forming galaxies.  Nevertheless, we explore the effect of TNG's feedback implementation on the growth of the galaxy core.

We examine the effects of BH feedback in \fig{bhmdot} which shows the steepness of the core growth in $\sigtwo$-$\Ms$ over 2 Gyr (left) and 4 Gyr (right) as a function of the {instantaneous} BH mass accretion rate $\bhmdot$, normalized by $\Ms$, at the start (blue) and end (orange) of the evolution.  The BH accretion rate measured at the start of a 2 Gyr evolutionary track predicts the future steepness of the core growth in that track (blue curve of the left panel).  However, the correlation weakens for longer time intervals, almost disappearing for 4 Gyr long evolution (blue curve of the right panel).  On the other hand, the orange curves show that the BH accretion rate measured at the end of the evolution is a stronger indicator of past core growth: higher accretion rates indicate that the galaxy grew its core density quickly relative to its total mass.  This correlation also weakens for longer time intervals, but does not disappear after 4 Gyr.  Therefore we conclude that {\it the BH accretion rate is an indicator of past core growth} but not a predictor of future core growth. The results shown in \fig{bhmdot} support the scenario in which gas inflows build the core and feed the BH.

\begin{figure*}
\includegraphics[width=0.45\linewidth]{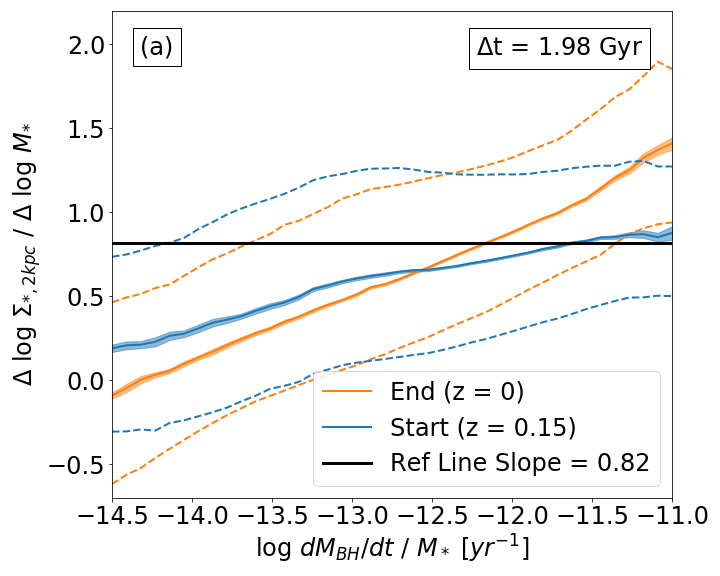}
\includegraphics[width=0.45\linewidth]{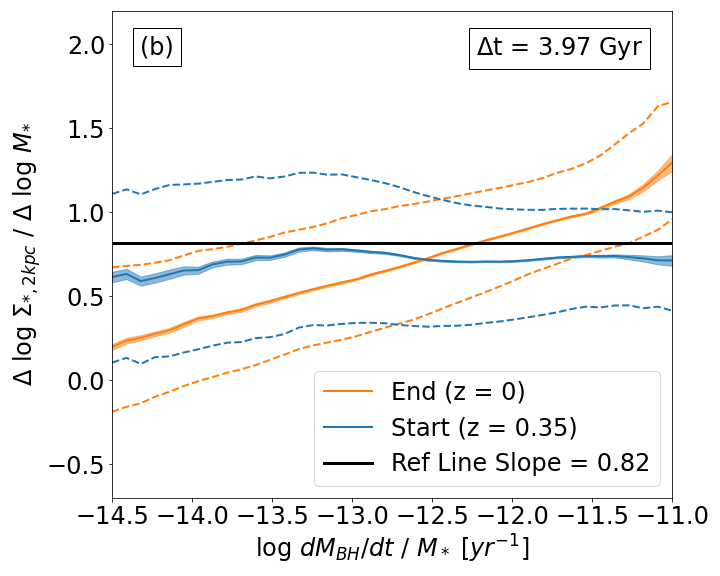}
\caption{The running median values of the slope of the evolutionary trajectory in the $\sigtwo$-$\Ms$ plane as a function of the normalized BH accrection rate at the start (blue) and end (orange) of a 1.98 Gyr (left) and 3.97 Gyr (right) trajectory.  Quartiles are indicated by the dashed lines.
The black reference line shows the slope of the {least-squares fit to the} $\sigtwo$-$\Ms$ relation at $z=0$.
{The instantaneous BH accretion rates are an indicator of past core growth over at least 4 Gyr.  However the BH accretion rate is not a strong predictor of future core growth, with any predictive power disappearing after 4 Gyr.} }
\label{bhmdot}
\end{figure*}

Perhaps it is not unexpected that the BH accretion rate does not predict future core growth since it is an instantaneous quantity that varies considerably on rapid timescales.  \cite{Terrazas2020} found that BH feedback is best seen as a cumulative effect, particularly when studying galaxy quenching.  Therefore, we also investigated whether the evolution of the core depends on $\MBH/\Ms$\footnote{
   $\MBH/\Ms$ is physically similar to the quantity $\int \dot{E}_{\rm kinetic} {\rm d}t / E_{\rm bind}$, i.e., the integral of the kinetic energy injected by the BH divided by the binding energy of the gas, which was found by \cite{Terrazas2020} to be the critical quantity for galaxy quenching.  We have found as yet unexplained problems with the binding energies for $\sim$ 15\% of the galaxies (confirmed by B. Terrazas, private communication), and so opted to use the simpler $\MBH/\Ms$.}.   
In \fig{mbhms} we show the dependence of the slope of the $\sigtwo$-$\Ms$ evolution over 2 Gyr on $\MBH/\Ms$ at the beginning (blue) and at the end (orange) of the evolution (the 4 Gyr interval is similar).  The blue curve shows that the normalized BH mass $\MBH/\Ms$ is somewhat predictive of subsequent core growth.  Galaxies with less massive BHs tend to grow their cores rapidly compared to their total mass growth.

\begin{figure}
\includegraphics[width=0.9\linewidth]{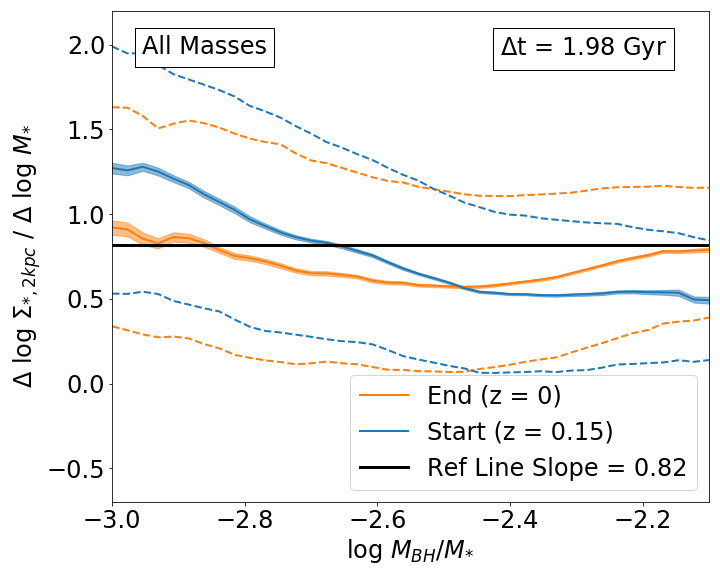}
\caption{The running median values of the slope of the evolutionary trajectory in the $\sigtwo$-$\Ms$ plane over the last 1.98 Gyr as a function of the normalized BH mass at the start (blue) and end (orange) of the trajectory.  Quartiles are indicated by the dashed lines.
The black reference line shows the slope of the {least-squares fit to the} $\sigtwo$-$\Ms$ relation at $z=0$.  
The BH mass $\sim 2$ Gyr ago predicts subsequent core growth for galaxies.}
\label{mbhms}
\end{figure}

{The predictive power of $\MBH/\Ms$ over subsequent core evolution is mass dependent.  In \fig{mbhms_splitmass} we show the same correlations as \fig{mbhms}, except split by galaxy mass (above and below $10^{10} \Msun$).  \fig{mbhms_splitmass} shows that the correlation between $\MBH/\Ms$ and the steepness of core growth is driven by the low-mass galaxies.  For galaxies more massive than $10^{10} \Msun$, BH mass no longer predicts how fast the core grows.  This mass dependence roughly corresponds to the mass scale above which the feedback mode switches from thermal to kinetic, the latter being the primary feedback that suppresses star formation.  \fig{bhmasses} shows the distribution of BHs and their cumulative kinetic feedback in the $\sigtwo$-$\Ms$ plane. Kinetic feedback dominates above the same characteristic $\Ms$ where steeper core evolutions cease (\mfigs{hists}{slopehist}).  At low masses, the thermal mode fails to prevent star formation in the core and $\sigtwo$ can grow rapidly compared to $\Ms$.}

\begin{figure*}
\includegraphics[width=0.45\linewidth]{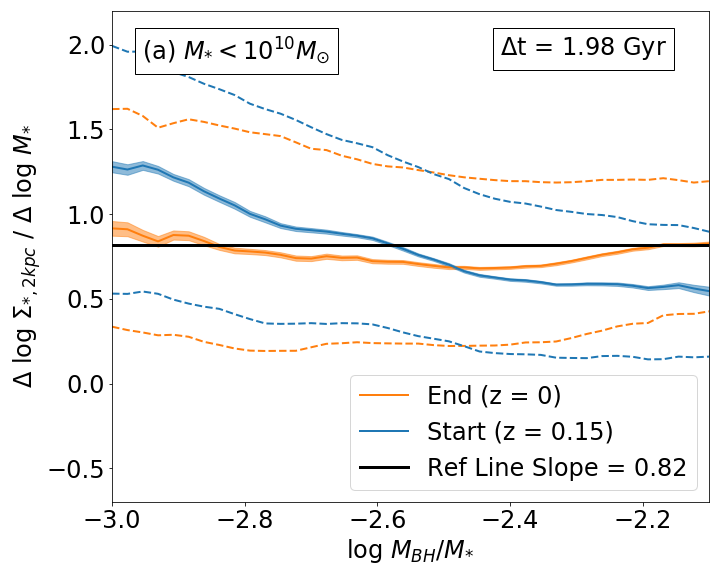}
\includegraphics[width=0.45\linewidth]{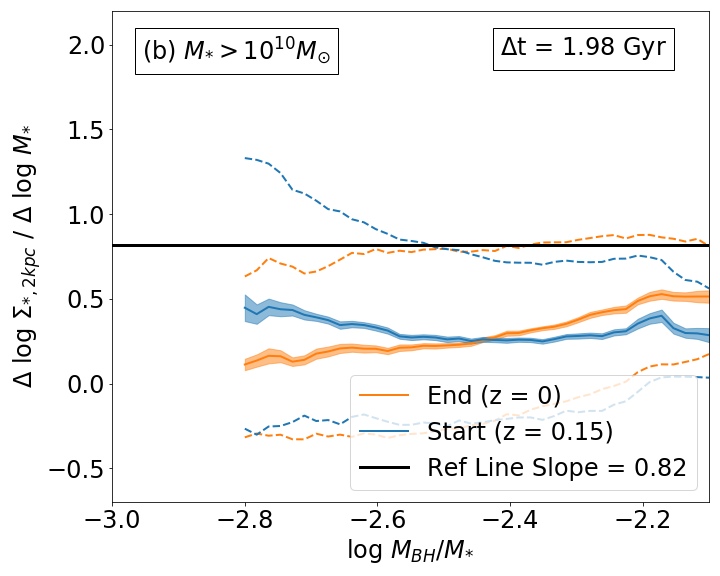} 
\caption{The running median values of the slope of the evolutionary trajectory in the $\sigtwo$-$\Ms$ plane over the last 1.98 Gyr as a function of the normalized BH mass at the start (blue) and end (orange) of the trajectory.  Quartiles are indicated by the dashed lines.  The left and right panels show galaxies below and above $\Ms = 10^{10} \Msun$.
The black reference line shows the slope of the {least-squares fit to the} $\sigtwo$-$\Ms$ relation at $z=0$.  
The correlation between BH mass and core growth is driven by the low mass galaxies.  Massive galaxies show very little correlation (where statistics are sufficient).
}
\label{mbhms_splitmass}
\end{figure*}

\begin{figure}
\includegraphics[width=1.1\linewidth]{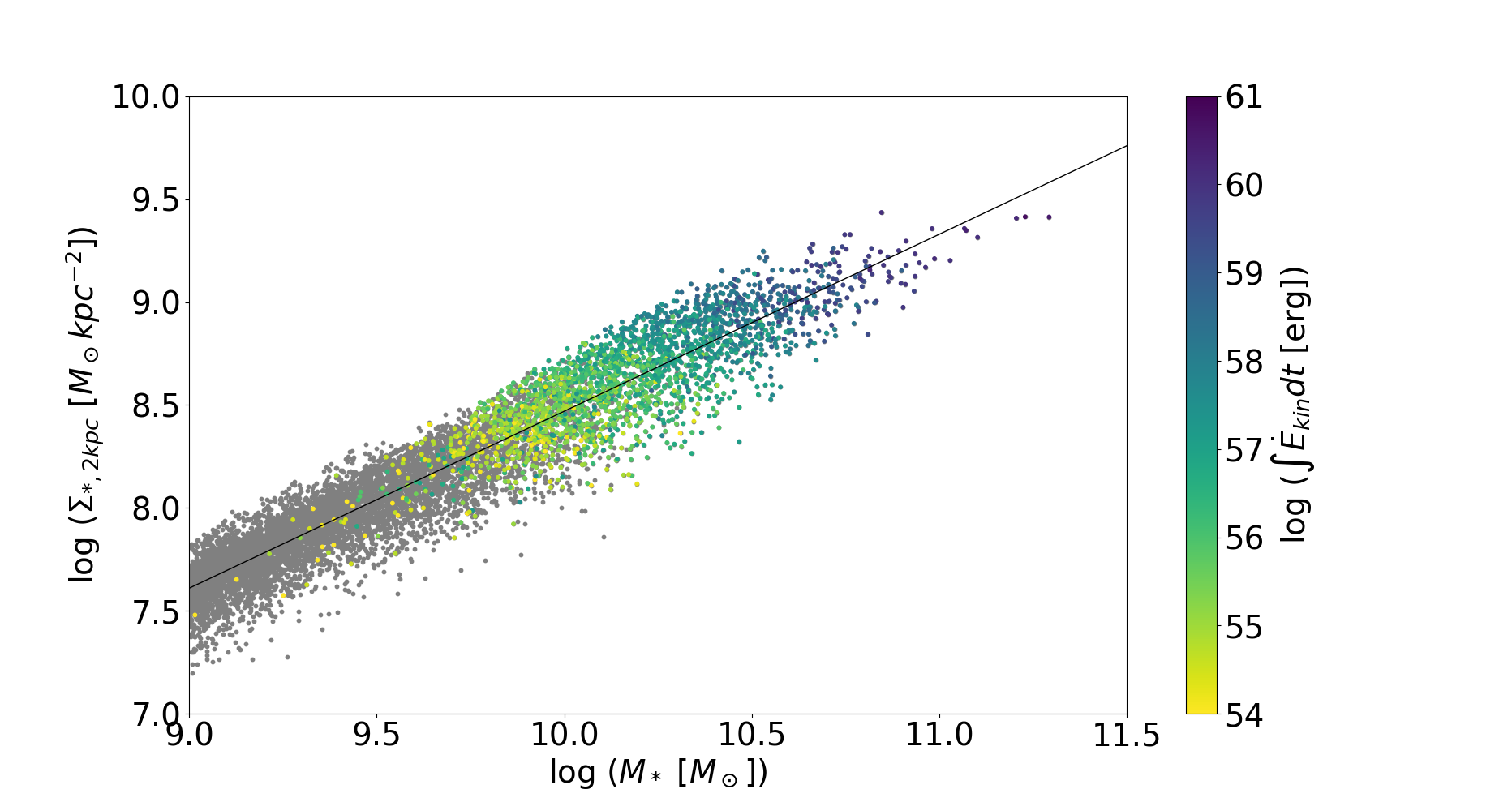}
\caption{Isolated SF subhaloes in the $\sigtwo$-$\Ms$ plane, coloured by $\int \dot{E}_{\rm kinetic} {\rm d}t$. Grey dots indicate that the subhaloes' BHs have not produced any feedback in kinetic mode. Note the mass dependence, with kinetic mode BHs becoming prolific above $\Ms \sim$ $10^{9.5}-10^{10} M_{\odot}$. }
\label{bhmasses}
\end{figure}

The results from \mfigs{bhmdot}{bhmasses} together suggest that although instantaneous BH feedback is a poor predictor of core growth, {\it the cumulative BH feedback history may be suppressing rapid core growth} seen in isolated star-forming galaxies below $\Ms \sim 10^{9.5}\Msun$. This may also explain the merger stellar mass effect seen above.

\subsubsection{Effects of Gas Angular Momentum}
\label{angmom}

Lastly, we investigate the effect of the angular momentum of the accreting gas on galaxy structural evolution.  The angular momentum properties of the gas (which correlate with the angular momentum of the stars and host halo) correlate with galaxy morphology \citep{Fall1983,Romanowsky2012,Zavala2016,Garrison-Kimmel2018}.  
Low-angular-momentum gas can reach the centre of the galaxy and build the core, whereas gas with high angular momentum likely contributes to star formation in the disc.

In order to study the effect of angular momentum on core growth, we calculate the specific angular momentum of gas crossing a shell of $R_{vir}/2$.  The infall time from this point is $\ltsima$ 1 Gyr. We select gas particles outside this shell with inwards radial velocity such that they would cross the shell within 100 Myr. We check that varying this period from 10 Myr to 1 Gyr has no significant impact on our results.

\fig{angmomcomp} shows the correlation between the evolutionary slope in the $\sigtwo$-$\Ms$ plane and the 
total specific angular momentum of gas particles at the start (blue) and end (orange) of 2 Gyr of evolution (4 Gyr is similar).
As expected, the blue curve shows that subhaloes whose infalling gas had low-angular momentum at the start of the evolution will experience steeper evolutionary slopes in the $\sigtwo$-$\Ms$ plane, and vice-versa.

The orange curve of \fig{angmomcomp} shows that the gas angular momentum at the end of the evolution also correlates with the steepness of core growth with similar correlation strength as the blue curve.  Given that the infalling gas has not yet reached the galaxy, there is no reason to expect a correlation between the end-point angular momentum and the galaxy's core growth, unless the accretion and its angular momentum are sustained over long periods ($>$ 2 Gyr) with lasting effect on core growth.  Indeed we confirm that the same correlation between core growth and the total specific angular momentum of the accreting gas (both before and after the evolution) is sustained for at least the last 5 Gyr. Therefore, {\it structural evolution, namely rapid core growth, is driven by the angular momentum of the accreting gas}.

\begin{figure}
\includegraphics[width=0.9\linewidth]{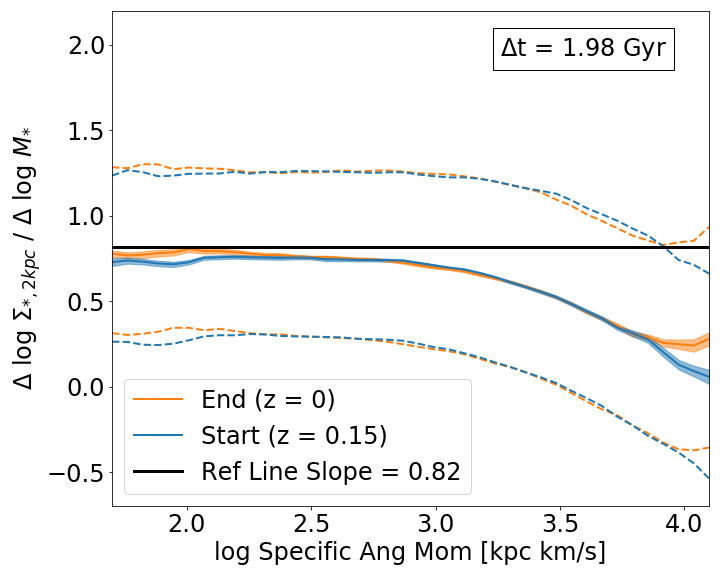}
\caption{The running median values of the slope of the evolutionary trajectory in the $\sigtwo$-$\Ms$ plane over the last 1.98 Gyr as a function of the specific angular momentum of accreting gas as measured at half the virial radius at the start (blue) and end (orange) of the trajectory.  Quartiles are indicated as dashed lines.  The black reference line show the slope of the {least-squares fit to the} $\sigtwo$-$\Ms$ relation at $z=0$.
Higher specific angular momentum of accreting gas $\sim 2$ Gyr ago predicts slower core growth compared to the total mass.  This correlation is maintained even when the specific angular momentum is measured at the end of the trajectory indicating that the link between angular momentum and core growth is long-lived.}
\label{angmomcomp} 
\end{figure}

\section{Discussion}
\label{discussion}

\subsection{Drivers of Galaxy Evolution on the $\sigtwo$-$\Ms$ Plane}
\label{drivers}

We have demonstrated that TNG100-1 {{qualitatively}} reproduces many observed structural properties, making it a useful tool to investigate the origin of the observed galaxy structure. Using TNG100-1, we are able to answer key questions pertaining to galaxy evolution, for example: why do galaxies with dense cores also have centrally concentrated star-formation?  What is the history of structural evolution for isolated star-forming galaxies, and what drives the evolution of the core?  These questions were ultimately motivated by the existence of the morphology-quiescence relation, where the density of the core is strongly correlated with quiescence.  We defer a treatment of quiescent galaxies to a later study, while we {first} attempt to understand the growth of the core in star-forming galaxies.

In \secref{toomredisc} we demonstrated that disc instabilities are not a major driver of structural evolution in isolated galaxies at low-$z$.  There is little correlation between Toomre Q and core growth over the 5 Gyr prior to $z=0$.  In fact, we have found that Toomre Q {\it responds} to the structural evolution, such that rapid core growth results in more stable disc.  Our findings are consistent with the scenario where dispersion supported central spheroids stabilize galactic discs \citep{Martig2009,Gensior2020}. 

Our findings that disc instabilities are unimportant for low-$z$ core building is in contrast to what is seen in hydrodynamical simulations at high-$z$ ($z\sim 2$); \citep{Zolotov2015}.  At cosmic ``high-noon'', discs are violently perturbed by intense instreaming, which include minor mergers.  Minor mergers are not very different from violent disc instabilities since minor mergers occur frequently relative to the disc orbital time.  At low-$z$, mergers are less frequent, especially for isolated galaxies; 
{{we have shown here that although mergers tend to lead to core build-up, the effect decreases as galaxy mass increases and gas fraction decreases. At the highest masses, dry mergers actually have the effect of diffusing the core stellar density.}}
This could be due to resolution effects, but could also be due to the fact that in TNG, BH feedback in low-mass galaxies is in the thermal mode which may be ineffective at removing star-forming gas from the core. However, in galaxies more massive $\Ms \sim 10^{10} \Msun$, the low-accretion kinetic mode {{is more effective at preventing mergers from funneling gas to the core.}}

{{We find that kinetic BH feedback, particularly at $\Ms \gtsima 10^{9.5} \Msun$, prevents steep core growth. \cite{Choi2018} recently found similar results in a set of zoom-in simulations using mixed kinetic and thermal AGN feedback, in particular for quenched galaxies. They found that the presence of the AGN halts $\sigone$ growth as galaxies quench. Further, a combination of stellar death, adiabatic expansion following gas expulsion, and binary BH scouring results in $\sigone$ decreasing over long times (as they call ``puffing-up'' the core).}}

While BH feedback appears to prevent the most rapid core growth above $\Ms \sim 10^{9.5} \Msun$, the variation in steepness of core growth between subhaloes is governed by the angular momentum of the accreting gas.
In the classical picture of galaxy formation, gas is accreted onto the galaxy in a ``hot'' mode (cooling from the shock-heated halo), and in a ``cold'' mode (dense and filamemtary that does not shock-heat), but the angular momentum of both modes should be the same as the halo (eg. \citealp{Fall1980}).  The angular momentum of the accretion (and thus also of the halo) correlates with galaxy morphology \citep{Romanowsky2012,Teklu2015,Zavala2016,Garrison-Kimmel2018}, and by extension with $\sigtwo$. However, a more complex and nuanced picture of the relation between angular momentum and the galaxy has {also emerged from recent work}. For example, cold filaments carrying high angular momentum are able to penetrate the hot halo without mixing with the low-angular momentum halo gas, so the accreting gas can have very different angular momentum from the halo {\citep{Stewart2017}}.  The angular momentum of the halo is the cumulative result of past accretion, whereas the cold gas in a halo is recent, entering the halo with higher specific angular momentum \citep{Danovich2015,Stewart2017}.  Thus it has been recognized that the angular momentum of accreting gas does not neccessarily determine galaxy morphology and core density \citep{Teklu2015}.

Instead, it is the alignment of the angular momentum of the gas with that of the galaxy that correlates most strongly with galaxy morphology \citep{Sales2012,Duckworth2020a}.  The hot low-angular-momentum gas cooling from halo often contributes significantly to the disc at later times, whereas the high-angular momentum cold accretion, which is not generally well-aligned with the halo or galaxy at late times, contributes to the spheroids.

In the present study, we have used $\sigtwo$ as a measure of ``morphology''.  $\sigtwo$ is tightly correlated with $\Ms$, but galaxies vary considerably in their evolutionary paths along the $\sigtwo$-$\Ms$ relation.  We have shown that the steepness of the evolutionary path in $\sigtwo$-$\Ms$, i.e., the rate of the growth of the core relative to the growth of the whole galaxy, correlates with the specific angular momentum of the accreting gas.  In our analysis, we have included all infalling gas, whether in the cold or hot mode.  We have also taken the vector sum of the angular momentum of all infalling gas particles, so that the alignment between cold and hot modes is taken into account. Non-aligned cold gas, even with high angular momentum, will lower the total specific angular momentum, resulting in steeper evolutionary tracks in $\sigtwo$-$\Ms$. We note that we see the same trends even when using the scalar sum of the angular momentum vectors which may suggest that non-aligned gas may not play an important role in core evolution.
Lastly, we note that the correlation between core growth and the total specific angular momentum of the accreting gas is long-lived ($\gtsima 5$ Gyr).  

Therefore, we have found that in the IllustrisTNG simulations, the growth of the galaxy core is related to both the mode of BH feedback (thermal vs kinetic), and to the angular momentum of accreting gas.  Above $\sim 10^{9.5}\Msun$, BH feedback slows core growth, but low-angular-momentum gas successfully overcomes the effects of BH feedback at all masses. In fact, the shallower trajectories of core growth occur at lower $\sigtwo$ (relative to $\Ms$ - see \mfigs{hists}{slopehist}), where the $\MBH$ is lower.

\cite{Woo2019} treated the shallower trajectory as the ``default'' growing mode (see also \citealp{Chen2020}).  We have seen that these shallow trajectories are due to a combination of high angular momentum accretion and BH feedback.  \cite{Woo2019} also interpreted the steeper trajectories as a ``compaction-like'' event that deviates from the default growing behaviour.  We have seen here that steeper tracks (above $\sim 10^{9.5}\Msun$) are caused by accretion of gas with lower total specific angular momentum, including non-aligned accretion with higher specific angular momentum.   
Our result that the correlation between low gas angular momentum and steep core build-up is long-lived perhaps implies that the steeper trajectories are not necessarily a special event, but rather a result of continuous accretion of low-angular-momentum or non-aligned material.  

Based on these findings, we propose a new cartoon for galaxy evolution in the $\sign$-$\Ms$ plane (\fig{cartoon4}). The slopes are shallower than proposed in the cartoon by \cite{Woo2019} (see their Fig. 2), and include a range of slopes that depend on position in the diagram rather than the two-track evolution they proposed.  Nevertheless, the salient point of their proposal is confirmed in TNG: {galaxies with diffuse cores at $z=0$ grew their total mass faster than their cores in the past, while those with dense cores by $z=0$ grew} their cores quickly relative to their total mass. Subhaloes with dense cores mostly grew parallel to the {fit} line (which is steep relative to the median evolutionary slope), and subhaloes with more diffuse cores relative to their mass mostly grew along shallower evolutionary tracks.

\begin{figure}
 \includegraphics[width=0.9\linewidth]{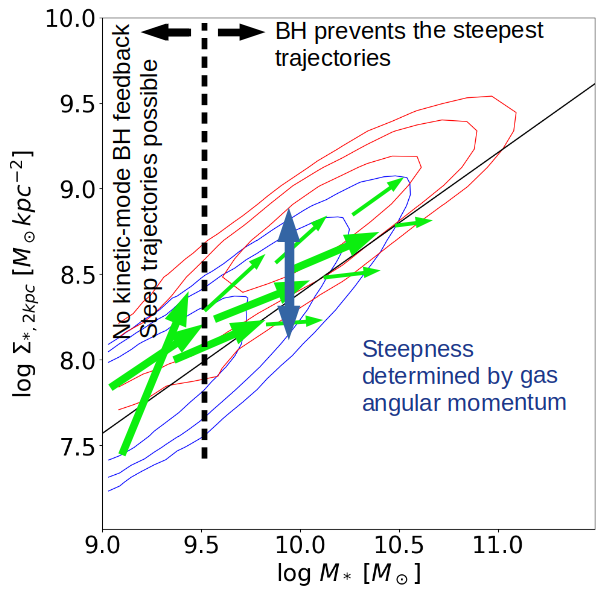}
 \caption{A summary of the evolution of {isolated} galaxies in the core-density $\sigtwo$ vs stellar mass $\Ms$ plane and the drivers of core growth in the IllustrisTNG simulation, {{at $z\sim0$}}. {Green arrows indicate schematic trajectories for individual galaxies. Contours represent number density of SF (blue) and quiescent (red) galaxies in SDSS ({reproducing Fig. 2 of WE19, except with $\sigtwo$ instead of $\sigone$}). In TNG, galaxies at low mass (left of the dashed black line) have low mass BHs which do not produce kinetic feedback, allowing steep core growth relative to total $\Ms$. At all masses, {galaxies with higher $\sigtwo$ by $z=0$ experienced steeper core growth in the past.  Furthermore} the steepness of a galaxy's trajectory in $\sigtwo$-$\Ms$ is determined primarily by the angular momentum of the infalling gas from the halo.} {{The evolutionary pathways diverge slightly, resulting in a marginally wider distribution over time (\fig{subanim}). Quenching removes galaxies from the upper envelope of the SF relation, potentially keeping the relation relatively tight. We stress that this cartoon represents TNG behaviour at low redshift, as we have not investigated trajectories or drivers earlier than $z=0.5$.}}}
 \label{cartoon4}
\end{figure}

\subsection{Implications for the Evolution of the $\sigtwo$-$\Ms$ Relation}

The zero-point of the $\sigone$-$\Ms$ relation at $\Ms=10^{10.5}\Msun$ has been observed to increase with $z$ until $z\sim 2$ \citep{Barro2017,Mosleh2017}. {We confirm a similar evolution of the relation in TNG} 
(\fig{subanim}). However the individual galaxies evolve upwards and to the right in this diagram (they are always growing both their total masses and their cores).  If the evolutionary tracks are shallower than the main relation, the zero-point decreases: perhaps more accurately, the main relation evolves to the right (see \citealp{Chen2020}).  What is driving the shallower trajectories, and thus the observed zero-point evolution?  We have found that they are the result of both BH feedback and accreting gas with high angular momentum.  Without the accumulated feedback of the massive BHs, even high-angular-momentum accretion will eventually feed core growth.  However even with massive BHs, high-angular-momentum accretion is required to produce shallower trajectories of core growth.

Although the observed zero-point evolution of the $\sigtwo$-$\Ms$ relation is qualitatively reproduced in TNG, the slope of the relation evolves significantly in TNG (\ie, the relation is steeper with $z$), contrary to the remarkably constant slope in the observed universe \citep{Saracco2012a,Tacchella2015a,Barro2017,Mosleh2017,Tacchella2017}. Since galaxies growing in the steeper mode of core evolution have average slopes that are comparable to the slope of the relation, this may imply that too many galaxies at high $z$ in TNG are in the steeper mode of core growth.  In other words, either BH feedback is not effective enough in tempering core growth at high $z$, or the accreting gas falls in with angular momentum that is too low at high $z$.  Given the long-lived and slowly-evolving nature of cosmic inflows \citep{Stewart2017}, and given the neccessarily simplistic nature of the sub-grid implementation of BH feedback in large simulations, we suspect that it is the BH feedback that is to be blame, i.e., the feedback is not effective enough at high $z$.

Alternatively, or perhaps in addition to having too many rapidly growing cores at high $z$, the evolving slope of the $\sigtwo$-$\Ms$ relation may be due to TNG having too many galaxies at low $z$ on the shallow mode of core growth. 
In other words, either BH feedback is overly effective at stunting core growth at low $z$, or the accreting gas falls in with angular momentum that is too high. {Similarly, if BHs are too effective at low $z$, they may be driving many of the steeply evolving galaxies into quiescence, removing too many with high $\sigtwo$ from the $\sigtwo-\Ms$ relation for SF galaxies.} Again, we suspect that the BH feedback implementation is the culprit, rather than the gas accretion.

The overall implications of our findings, as well as the relatively tight scatter in the observed $\sigtwo$-$\Ms$ relation, imply that the BH feedback and the angular momentum of the accreting gas conspire to regulate the growth of galaxy cores in relation to the growth of their total stellar mass.  BH feedback and gas accretion seem to act in concert such that the slope of the $\sigtwo$-$\Ms$ remains constant while the zero-point decreases over time.

\subsection{Caveats}
\label{caveats}

We note that all of our findings regarding BH feedback come with the caveat that in TNG, black holes in star-forming galaxies are too massive by almost 2 orders of magnitude compared with observations.  Although the feedback implementation in TNG successfully reproduces the observed BH masses in the quiescent population, BHs grow too quickly in star-forming galaxies ({eg. see Fig. 7 in \citealp{Terrazas2020}).  It remains to be seen whether the feedback physics in TNG could be scaled to produce lower BH masses in star-forming galaxies while still successfully quenching galaxies. 

We note that our interpretations are based on the assumption that the formation of a dense core must be caused by an inflow of gas, fuelling star formation in the galactic core. The other possibility is that stars migrate inwards. However, apart from mergers or other galaxy interactions, we see no mechanism that could cause this  {{on a  significant scale (although slow outward migration of stars can occur as shown by \cite{Choi2018})}}. We have shown that mergers are rare in our sample, and since we have selected only isolated galaxies, we expect other interactions are also rare.

We also note that studying the inner 1-2 kpc of subhaloes is pushing the resolution limit of TNG100-1. {In particular, our examination of disc instabilities was limited to a scale of 0.74 kpc. Most SF regions are much smaller ($\sim$10-100pc). We acknowledge that the resolution of TNG100-1 does not allow us to probe disc instabilities down to these scales.} In this work, we present $\sigtwo$ instead of $\sigone$ due to resolution concerns. {We did run all analyses for $\sigone$ in addition to $\sigtwo$ and found all trends were the same. In TNG, subhaloes exhibit more spread in $\sigone$ than $\sigtwo$, as do galaxies in SDSS and MaNGA. Low mass subhaloes $< 10^{9.5} \Msun$ exhibit even steeper growth in $\sigone$-$\Ms$ than in $\sigtwo$-$\Ms$.}

To check for convergence and any effects of resolution, we also ran our analyses on TNG100-2. TNG100-2 has roughly half the spatial resolution of TNG100-1. For example, the gravitational softening length for collisionless particles in TNG100-1 is 0.74 kpc while in TNG100-2 it is 1.48 kpc \citep{Nelson2019}. In general, TNG100-2 produces a similar distribution of SF, intermediate, and quiescent subhaloes as \fig{qtsfcent}, with better agreement at $\Ms > 10^{10} \Msun$. The three sequences occupy about the same ranges of $\Ms$ as TNG100-1. However, the sequences in TNG100-2 all sit about {0.1-0.2 dex lower in $\sigtwo$ and about 0.3-0.4 dex lower in $\sigone$.}

Although we must conclude that TNG100-1 may not be fully converged in $\sigtwo$ or $\sigone$, the trends seen in \threefigs{agncompare}{radgrads1}{radgrads2} are all present in both TNG100-2 and TNG100-1. We also found that subhalo histories were steeper for galaxies that end up with higher $\sigtwo$ by $z=0$, and vice-versa (as seen in \fig{hists}), in both simulation runs. TNG100-2, however, does not produce the clear limit at $\sim 10^{9.5} M_{\odot}$ for steep evolutionary tracks. In terms of correlating slopes of the evolutionary trajectories, TNG100-2 does produce roughly the same trends {seen in \fig{slopesfrslope} (sSFR gradient), \fig{toomrecorrelate} (Toomre Q), and \fig{bhmdot} (BH accretion rate). However the trends seen in \fig{mbhms} (BH mass) only agree above $\Ms > 10^{10} \Msun$. The correlation seen in  \fig{angmomcomp} (angular momentum) is not present in TNG100-2, although we note that the mass resolution of gas particles is an order of magnitude worse than TNG100-1. 

We acknowledge that the results presented may therefore not be fully converged. We look forward to the public release of TNG50 to check our results at higher resolution.} We present details of our convergence tests in Appendix A.

\section{Summary}
\label{summary}

We use the public data release of Illustris TNG to investigate the recent morphological evolution of isolated galaxies. Specifically we consider the {evolution in} core stellar density ($\sigtwo$) {relative to the evolution in} total stellar mass ($\Ms$). Our findings regarding the $\sigtwo$-$\Ms$ can be summarized as follows:

\begin{enumerate}
 \item {{Qualitatively,}} Illustris TNG produces the observed $\sigtwo$-$\Ms$ relations for isolated galaxies including: the star forming and quiescent sequences {(\twofigs{qtsfcent}{subanim})}, AGN distribution {(\fig{agncompare})}, {and} gradients of star formation rate {(\fig{radgrads1})}, stellar age {(\fig{radgrads2})}, { and gas phase metallicity {(\fig{radgrads3})}.}
 \item {
 {The progenitors of galaxies with diffuse cores at z = 0 (at fixed $\Ms$) evolve along shallow trajectories in the $\sigtwo$-$\Ms$ plane. Conversely, the progenitors of galaxies with dense cores at z = 0 evolve along steeper evolutionary tracks {(\twofigs{hists}{slopehist})}.}
 65\% of galaxies with $\Ms > 10^{9.5} M_{\odot}$ have shallower trajectories than the main distribution in $\sigtwo$-$\Ms$. Those with dense cores evolve roughly parallel to the main distribution of galaxies.
 However, low mass galaxies ($< 10^{9.5} M_{\odot}$) mostly evolve steeply in the $\sigtwo$-$\Ms$ plane, {likely due to the inability of BH feedback to prevent star-formation in the core.}}
\end{enumerate}

{We investigate possible drivers of steep core growth, namely high central sSFR, merging, disc instabilities, weak BH feedback and low angular momentum of accretion.  We find that:}

\begin{enumerate}
 \setcounter{enumi}{2}

  \item {A galaxy's instantaneous sSFR gradient is an indication of its past core evolution (over the last $\sim 4$ Gyr), such that galaxies with more centrally concentrated sSFR tended to have more rapid core growth relative to the growth of their overall mass.  The sSFR gradient in the past is not a major predictor of future core growth.  (\fig{slopesfrslope}).}
 \item Major mergers {are} rare in isolated galaxies since z = 0.5 (10\% of our sample). 
{{Mergers cause steeper core growth, but the effect decreases with increasing mass and decreasing gas fraction (\fig{mergereffect}). Additionally, most mergers at $\Ms > 10^{10}\Msun$ are dry.}} 
 \item Disc instability at a resolution scale of 0.74 kpc (as measured by the fraction of mass with Toomre $Q < 2$) does not predict the evolution of the core {(\fig{toomrecorrelate})}.  Rather, steeper core growth in the past predicts more stable discs today, indicating that dense cores have a stabilising effect on the disc.
 \item {High instantaneous BH accretion rate is an indicator of rapid core growth over the past 4 Gyr (relative to the growth of $\Ms$), but the BH accretion rate in the past is not a major predictor of future core growth (\fig{bhmdot}).  However the cumulative BH feedback history as measured by $\MBH/\Ms$ does predict future core growth for low-mass ($\Ms \ltsima 10^{9.5}\Msun$) galaxies {(\twofigs{mbhms}{mbhms_splitmass})} in the sense that lower $\MBH/\Ms$ predicts rapid core growth.} Low-mass galaxies have low $\MBH$, therefore they tend to feedback in the thermal AGN mode, which couples less effectively with the surrounding gas than kinetic mode. We interpret this as the cause of steeper core growth (\fig{bhmasses}).
 
 For more massive galaxies ($\Ms \gtsima 10^{9.5} \Msun$), trajectories in $\sigtwo$-$\Ms$ are on average shallower than for the less massive galaxies, suggesting that kinetic feedback was suppressing star-formation within the feedback region of 2-3 kpc. However, we found no correlation between $\MBH$ and core growth for these massive galaxies.
 \item The total specific angular momentum of the accreting gas in the halo predicts the steepness of core growth {(\fig{angmomcomp})}. Galaxies accreting high angular momentum gas tend to evolve along shallower slopes in the $\sigtwo$-$\Ms$ plane.  Furthermore this correlation persists whether measuring the angular momentum at the beginning of the evolutionary track or at the end of the track, even after 5 Gyr of evolution.  This indicates that the connection between the angular momentum of the accreting gas and the steepness of core growth is long-lived.
\end{enumerate}

{The growth of galaxy cores in the IllustrisTNG model is complex and is related to at least two physical drivers (the angular momentum of gas accretion and the BH feedback).}  It is entirely plausible that other causal factors that we have not considered are also at play, {especially given the large scatter in the correlations presented here}. 

\section*{Data Availability}

Data from the IllustrisTNG simulations are publicly available at \url{www.tng-project.org}. Data specific to this paper are available on request from the corresponding author.

\section*{Acknowledgements}
We thank the anonymous referee for the insightful comments. We acknowledge the helpful and stimulating discussions with 
Andi Burkert, Sandy Faber, Nir Mandelker, Dave Patton, Salvatore Quai, Sandro Tacchella, Bryan Terrazas and Elad Zinger. SLE gratefully acknowledges the receipt of an NSERC Discovery Grant. MHH acknowledges support from the William and Caroline Herschel Post-doctoral Fellowship fund, and the Vanier Canada Graduate Scholarship.

The simulations of the IllustrisTNG project used in this work were undertaken with compute time awarded by the Gauss Centre for Supercomputing (GCS) under GCS Large-Scale Projects GCS-ILLU and GCS-DWAR on the GCS share of the supercomputer Hazel Hen at the High Performance Computing Center Stuttgart (HLRS), as well as on the machines of the Max Planck Computing and Data Facility (MPCDF) in Garching, Germany.

This research made use of Astropy,\footnote{http://www.astropy.org} a community-developed core Python package for Astronomy \citep{Astropy2013,Astropy2018}. 
This research also made use of the computation resources provided by Westgrid (www.westgrid.ca) and Compute Canada (www.computecanada.ca).  

Funding for the Sloan Digital Sky Survey IV has been provided by the Alfred P. Sloan Foundation, the U.S. Department of Energy Office of Science, and the Participating Institutions. SDSS acknowledges support and resources from the Center for High-Performance Computing at the University of Utah. The SDSS web site is www.sdss.org.  SDSS is managed by the Astrophysical Research Consortium for the Participating Institutions of the SDSS Collaboration including the Brazilian Participation Group, the Carnegie Institution for Science, Carnegie Mellon University, the Chilean Participation Group, the French Participation Group, Harvard-Smithsonian Center for Astrophysics, Instituto de Astrofísica de Canarias, The Johns Hopkins University, Kavli Institute for the Physics and Mathematics of the Universe (IPMU) / University of Tokyo, the Korean Participation Group, Lawrence Berkeley National Laboratory, Leibniz Institut für Astrophysik Potsdam (AIP), Max-Planck-Institut für Astronomie (MPIA Heidelberg), Max-Planck-Institut für Astrophysik (MPA Garching), Max-Planck-Institut für Extraterrestrische Physik (MPE), National Astronomical Observatories of China, New Mexico State University, New York University, University of Notre Dame, Observatório Nacional / MCTI, The Ohio State University, Pennsylvania State University, Shanghai Astronomical Observatory, United Kingdom Participation Group, Universidad Nacional Autónoma de México, University of Arizona, University of Colorado Boulder, University of Oxford, University of Portsmouth, University of Utah, University of Virginia, University of Washington, University of Wisconsin, Vanderbilt University, and Yale University.

\bibliographystyle{mnras}
\bibliography{library}

\label{lastpage}

%% file: appendix.tex
 \renewcommand{\theequation}{A\arabic{equation}}
  \renewcommand{\thesection}{A\arabic{section}}
  \renewcommand{\thefigure}{A\arabic{figure}}
    \setcounter{equation}{0}    \setcounter{section}{0}    \setcounter{figure}{0}  
  \section*{APPENDIX A: Convergence}
  \label{appendix}

In this appendix we provide details of our numerical convergence tests. To check for convergence, we qualitatively compare our results using TNG100-1 to TNG100-2. Resolution and convergence in complex non-linear simulations are not straightforward. TNG100-2 has roughly half the spatial resolution of TNG100-1 (e.g. $z=0$ gravitational softening length of the collisionless component is 1.48 kpc vs 0.74 kpc) and roughly an order of magnitude lower mass resolution (e.g. dark matter particle mass 5.9 x $10^7 \Msun$ vs 7.5 x $10^6 \Msun$ and target baryon particle mass 1.1 x $10^7 \Msun$ vs 1.4 x $10^6 \Msun$) \citep{Nelson2019}. The size of Voronoi gas cells varies by about 4 orders of magnitude within each simulation. For example, in TNG100-1, star-forming gas cells can be as small as 20 pc while gas cells in the intergalactic medium can be as large as 200 kpc (see \cite{Nelson2019}, their Figure 2). 

\begin{figure}
\includegraphics[width=0.9\linewidth]{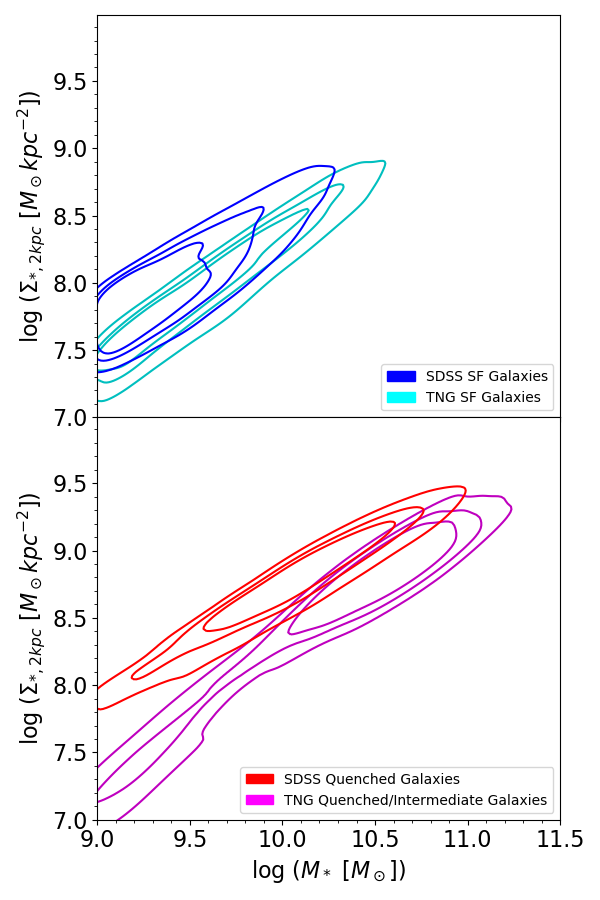}
\caption{\small The distribution of isolated galaxies in central stellar mass density ($\sigtwo$) vs total stellar mass ($\Ms$) in the SDSS and TNG100-2, split into SF (top) and quenched/intermediate galaxies (bottom). Contour lines represent the same normalized number densities of galaxies for both SDSS and TNG, and are drawn at {{PDF = 0.1, 0.3, and 0.5}}. SDSS galaxy densities are corrected for luminosity. TNG100-2 produces roughly the same distribution in $\Ms$ as TNG100-1 but with lower $\sigtwo$ by about 0.1-0.2 dex.}
\label{qtsfcent2}
\end{figure}

\begin{figure}
\includegraphics[width=0.9\linewidth]{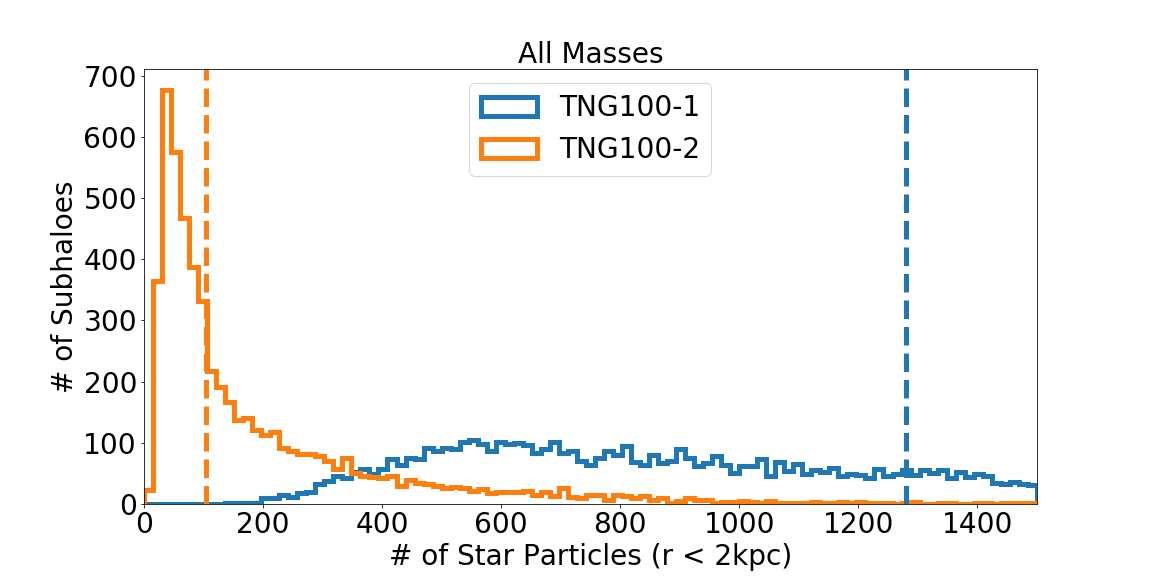}
\caption{\small Number of star particles inside 2 kpc for each simulation run. Dashed vertical lines show the median values in each case. 90\% of subhaloes in TNG100-2 have $<$ 500 particles inside 2 kpc, so it is not surprising that $\sigtwo$ is not fully converged between simulation runs.}
\label{particles_all}
\end{figure}

\fig{qtsfcent2} shows the equivalent of \fig{qtsfcent} but for TNG100-2. $\sigtwo$ values for SF subhaloes in TNG100-2 are $\sim$0.1-0.2 dex lower than in TNG100-1. Although not shown, we similarly find that TNG100-2 produces $\sigone$ values $\sim$0.3-0.4 dex lower than TNG100-1. At higher $\Ms$, the difference is less, and indeed the slope of a least squares fit to the TNG100-2 SF subhaloes at $z=0$ is 0.96 (compared with 0.86 for TNG100-1). To explore this further, \fig{particles_all} shows the number of star particles within 2 kpc for both simulation runs, for our entire sample of isolated SF subhaloes with $\Ms > 10^9 \Msun$. On average, the stellar component inside 2 kpc is not very well resolved in TNG100-2, with a median value of only 104 particles compared with 1280 for TNG100-1. About 90\% of subhaloes in TNG100-2 have $<$ 500 stellar particles inside of 2 kpc. If we restrict our sample to $\Ms > 10^{10}\Msun$ (not shown for brevity), TNG100-2 is more reasonable, with roughly half of all subhaloes having at least 500 particles inside 2 kpc. This is consistent with \fig{qtsfcent2}. Higher mass subhaloes are closer to convergence, so the values of $\sigtwo$ are closer between runs, which also results in a higher slope of the least squares fit in TNG100-2. 

\begin{figure}
\includegraphics[width=0.9\linewidth]{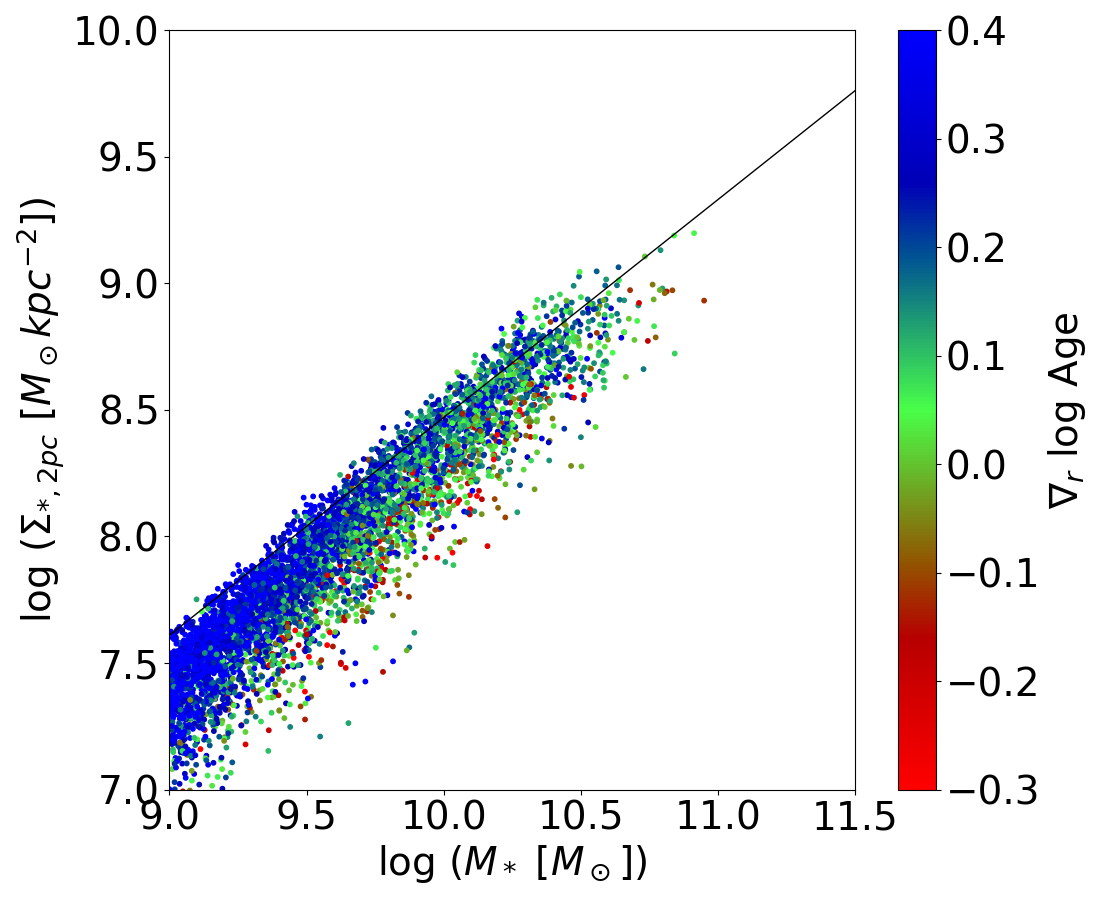}
\caption{The same as \fig{radgrads1} but for TNG100-2. Gradients of stellar age are plotted as a function of position in the {$\sigtwo$}-$\Ms$ plane. The line shown, for comparison, is the least squares fit to the TNG100-1 SF subhaloes. The TNG100-2 subhaloes lie farther from the TNG100-1 line at low $\Ms$ as previously noted, resulting in a steeper distribution. Qualitatively, both simulation runs produce more positive age gradients at higher $\sigtwo$, in agreement with observations. The agreement is weaker at low $\Ms$ where the resolution of TNG100-2 is questionable.}
\label{ageslopes2}
\end{figure}

\fig{ageslopes2} shows the radial gradients of stellar age in the $\sigtwo$-$\Ms$ plane for TNG100-2 (equivalent to \fig{radgrads1}). Qualitatively, both simulation runs produce more positive age gradients at higher $\sigtwo$, which is in agreement with observations. In TNG100-2 the trend is weak at low $\Ms$. Although not shown, the equivalent plots for sSFR and gas metallicity gradients, as well as AGN strength, are also generally in agreement.

Although not shown, TNG100-2 roughly agrees with \ref{slopehist}.
As in TNG100-1, subhaloes that have higher $\sigtwo$ at $z=0$ evolved along steeper trajectories in $\sigtwo$-$\Ms$, and vice-versa. However, there is no clear difference in evolutionary trajectories based on $\Ms$ at $10^{9.5} \Msun$.

\begin{figure}
\includegraphics[width=0.8\linewidth]{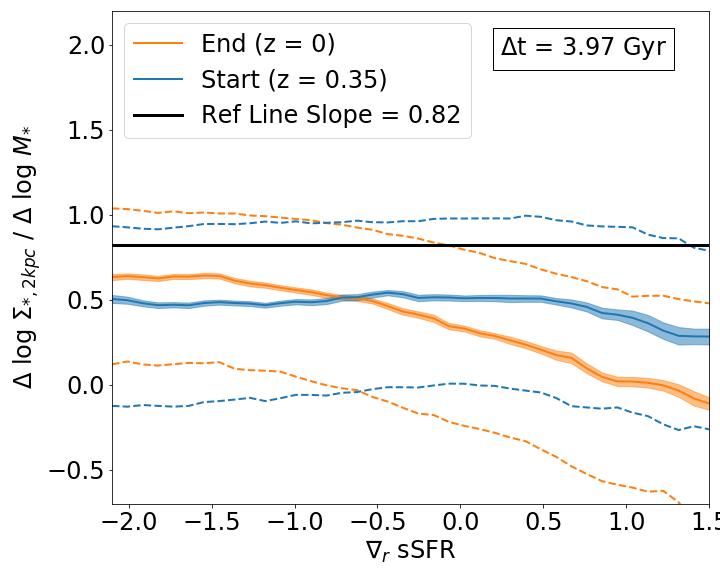}
\caption{Running median values of the slope of the evolutionary trajectory in the $\sigtwo$-$\Ms$ plane as a function of galaxy sSFR gradient at the start (blue) and end (orange) of a 4 Gyr trajectory, equivalent to \fig{slopesfrslope} (right hand panel) but for TNG100-2. TNG100-2 produces similar correlation to TNG100-1.}
\label{slopesfrslope2}
\end{figure}

\fig{slopesfrslope2} shows the correlation of evolutionary slopes with radial sSFR gradients in TNG100-2. TNG100-2 produces the same trends as TNG100-1 (\fig{slopesfrslope}, right hand panel). TNG100-2 also agrees for $\Delta$t = 2 Gyr (not shown), but with wider scatter. Analysis of several other drivers of core growth in TNG100-2 agrees roughly with TNG100-1: mergers, disc instabilities, and BH accretion rate. For brevity we do not show these here. Two drivers do not agree well: infalling gas angular momentum and $M_{BH}$.

\begin{figure}
\includegraphics[width=0.9\linewidth]{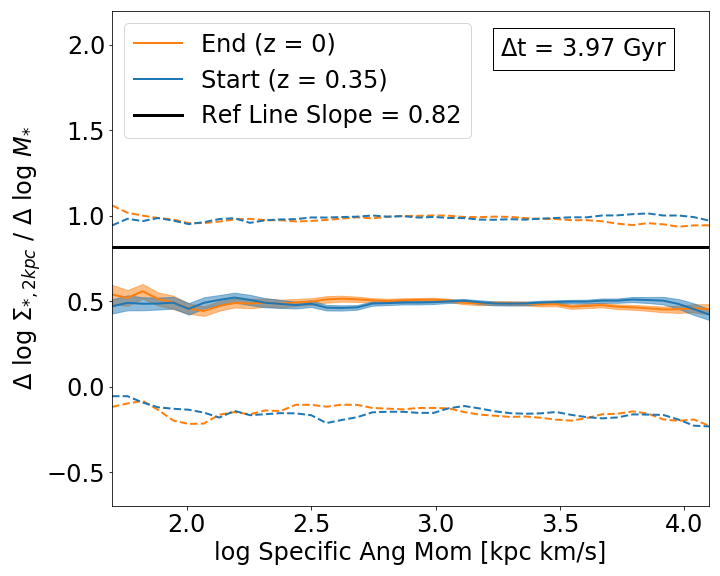}
\caption{The same as \fig{angmomcomp}, showing the running median values of the slope of the evolutionary trajectory in the $\sigtwo$-$\Ms$ plane over the last 4 Gyr as a function of the specific angular momentum of accreting gas as measured at half the virial radius at the start (blue) and end (orange) of the trajectory. There is no correlation in TNG100-2, which we suspect is due to the much lower mass resolution of gas particles compared to TNG100-1.}
\label{angmomcomp2}
\end{figure}

\begin{figure}
\includegraphics[width=0.9\linewidth]{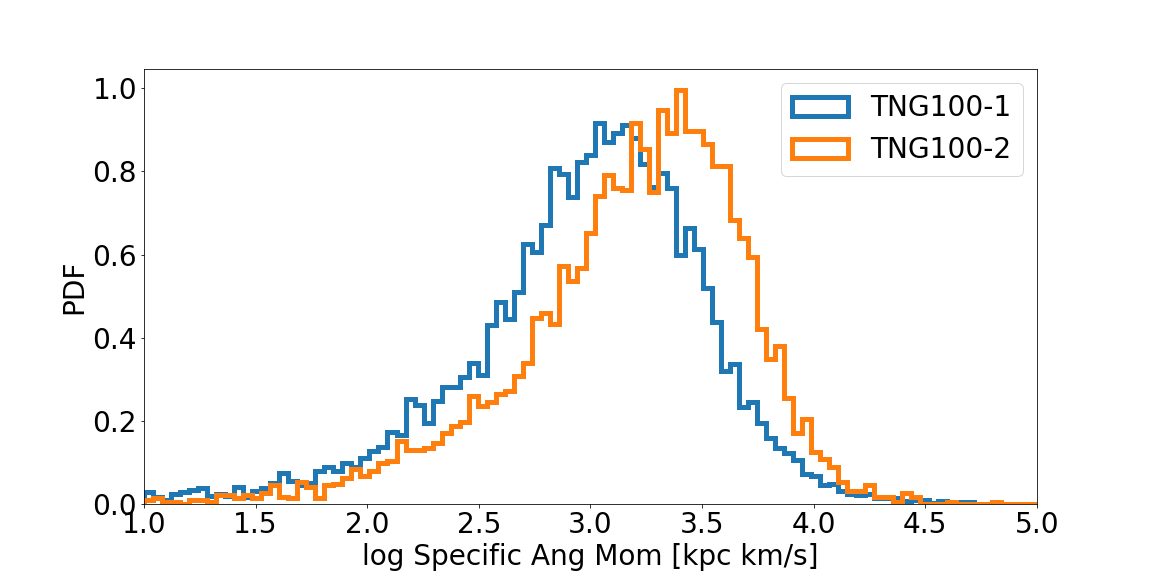}
\caption{Distribution of infalling gas specific angular momenta for all isolated SF subhaloes in both TNG100-1 (blue) and TNG100-2 (orange). The distributions are not converged between simulations, with the TNG100-2 distribution offset to higher specific angular momentum by about 0.3 dex.}
\label{angmomdist}
\end{figure}

First, we examine the effect of infalling gas angular momentum on core growth in TNG100-2 (\fig{angmomcomp2}, compare to \fig{angmomcomp} from TNG100-1). No correlation exists in TNG100-2. This is true regardless of subhalo $\Ms$ or $\Delta$t (not shown for brevity). However, since the mass resolution of gas particles in TNG100-2 is about an order of magnitude worse than TNG100-1, this is not entirely surprising. Indeed, \fig{angmomdist} presents the distribution of specific angular momentum for both simulation runs, showing that this quantity is not converged. We do not discount our results from TNG100-1, but acknowledge that we have not shown convergence of the angular momentum. We look forward to the public release of TNG50 to investigate further.

\begin{figure}
\includegraphics[width=0.9\linewidth]{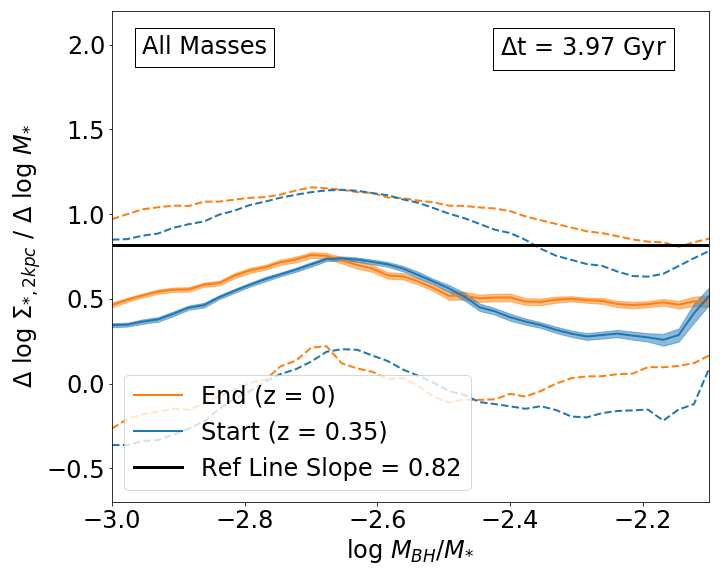}
\includegraphics[width=0.9\linewidth]{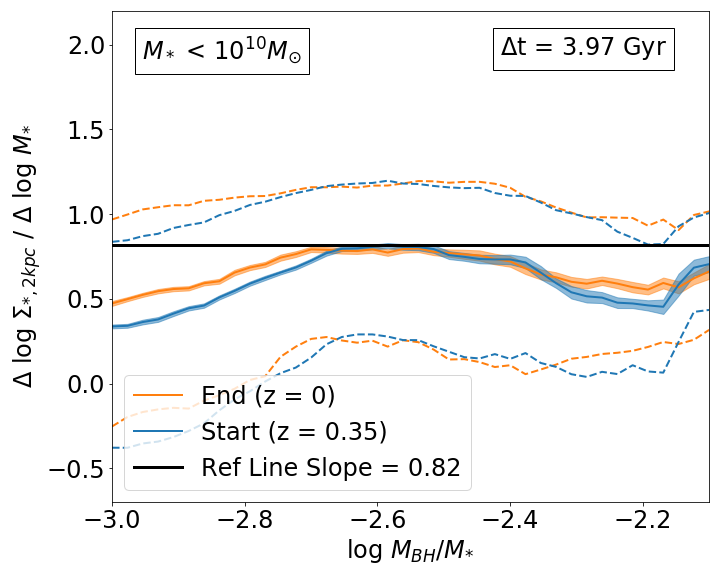}
\includegraphics[width=0.9\linewidth]{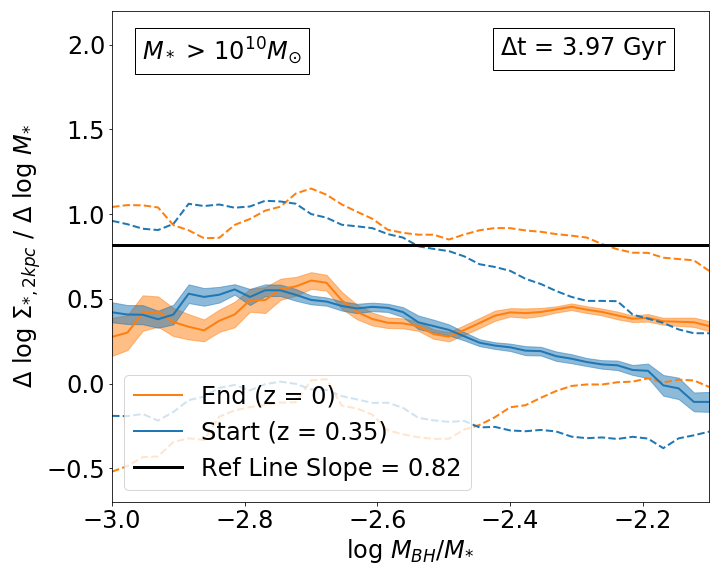}
\caption{The same as \fig{mbhms} (top panel) and \fig{mbhms_splitmass} (middle and bottom panels) but for TNG100-2. 
All panels show the running median values of the slope of the evolutionary trajectory in the $\sigtwo$-$\Ms$ plane over the last 4 Gyr as a function of the normalized BH mass at the start (blue) and end (orange) of the trajectory. Quartiles are indicated by the dashed lines. The black reference line shows the slope of the {least-squares fit to the} $\sigtwo$-$\Ms$ relation at $z=0$.  
The top panel shows all masses, while the bottom two panels split subhaloes at $\Ms = 10^{10} \Msun$. The correlation at high mass matches TNG100-1, indicating that the overall discrepancy is driven by low mass subhaloes in TNG100-2.}
\label{mbhms2}
\end{figure}

Lastly, we examine the effect of BH mass on core growth in TNG100-2. \fig{mbhms2} shows the correlation of evolutionary slope with $M_{BH}$ for TNG100-2, equivalent to \fig{mbhms} (top panel). The trends are not in agreement, particularly at low $M_{BH}$ where TNG100-1 predicts steep core growth. We split into 
mass bins (middle and bottom panels, compare to \fig{mbhms_splitmass}). TNG100-2 agrees at high mass, and therefore the overall discrepancy is driven by subhaloes with $\Ms < 10^{10} \Msun$. Note that again we use the higher $\Delta$t = 4 Gyr. $\Delta$t = 2 Gyr produces the same trends but with wider scatter.

